%% file: main.tex
\documentclass[sigplan,10pt]{acmart}

\acmSubmissionID{\#\paperid}
\renewcommand\footnotetextcopyrightpermission[1]{}
\usepackage[english]{babel}
\usepackage{blindtext}

\input{cmds.tex}

\input{macros.tex}

\setcopyright{none}

\acmDOI{}

\acmISBN{}

\acmConference[arXiv preprint]{}
\acmYear{2026}

\title{\Large \sysname: Enabling Collective Communication For Mixed-Vendor Heterogeneous Clusters}

\author{\Large Yuejie Wang}
\affiliation{
    \institution{\small Peking University}
    \country{}
}
\author{\Large Tao Chang}
\affiliation{
    \institution{\small Beijing Academy of Artificial Intelligence}
    \country{}
}
\author{\Large Yuanyuan Zhao}
\affiliation{
    \institution{\small Peking University}
    \country{}
}
\author{\Large Yulong Ao}
\affiliation{
    \institution{\small Beijing Academy of Artificial Intelligence}
    \country{}
}
\author{\Large Zeyu Gu}
\affiliation{
    \institution{\small Beijing Academy of Artificial Intelligence}
    \country{}
}
\author{\Large Zhiyu Li}
\affiliation{
    \institution{\small Beijing Academy of Artificial Intelligence}
    \country{}
}
\author{\Large Yanmin Jia}
\affiliation{
    \institution{\small Infrawaves}
    \country{}
}
\author{\Large Yan Zhang}
\affiliation{
    \institution{\small Infrawaves}
    \country{}
}
\author{\Large Mingjun Zhang}
\affiliation{
    \institution{\small Infrawaves}
    \country{}
}
\author{\Large He Liu}
\affiliation{
    \institution{\small Infrawaves}
    \country{}
}
\author{\Large Yongzhe He}
\affiliation{
    \institution{\small Institute of Computing Technology, Chinese Academy of Sciences}
    \country{}
}
\author{\Large Yonghua Lin}
\affiliation{
    \institution{\small Beijing Academy of Artificial Intelligence}
    \country{}
}
\author{\Large Guyue Liu}
\affiliation{
    \institution{\small Peking University}
    \country{}
}

\renewcommand{\shortauthors}{Wang et al.}
\renewcommand{\shorttitle}{\sysname}

\begin{abstract}

Training Large Language Models (LLMs) on heterogeneous clusters presents significant challenges for collective communication, as hardware from multiple vendors introduces diverse network and computational characteristics.
Existing collective communication frameworks (\eg NCCL, RCCL) designed for homogeneous environments fail to address mixed-hardware setups, while communication libraries with heterogeneous support (\eg Gloo, OpenMPI) incur heavy overhead in the data path.

This paper presents \sysname, a framework that enables heterogeneous collective communication by efficient P2P transport across heterogeneous devices (\eg GPUs), eliminating the host-device memory copy overhead while offloading the control to the CPUs.
For combining collectives (\eg AllReduce, ReduceScatter), \sysname introduces a border-communicator mechanism that achieves vendor independence by using the intrinsic reduction in the combining collectives in vendor collective communication libraries.
With efficient heterogeneous P2P transport and portable reduction mechanism, \sysname proposes a hierarchical topology abstraction for heterogeneous clusters, dissecting collective communication into cluster-level primitives that guarantee optimal cross-cluster data transfer volume and optimal bandwidth utilization.

We implement \sysname with 4 different vendor support and evaluate it in 4 heterogeneous settings with benchmarks and end-to-end LLM tasks.
Our evaluation shows that \sysname achieves $17 \sim 19 \times$ higher bandwidth than Gloo in heterogeneous communications, and speeds up end-to-end training by up to $16.9\%$ in the per-step-time.

\end{abstract}

\begin{document}
\maketitle

\input{1-intro}

\input{2-motivation}

\input{3-overview}
\input{4-design}
\input{5-impl}

\input{6-eval}

\input{7-conclusion}

\begin{acks}
This work was supported by National Natural Science Fund for the Excellent Young Scientists Fund Program (Overseas), and Peking University startup fund. This work was supported in part by the Beijing Major Science and Technology Project under Contract no. Z251100008125042. This work was supported by Beijing Academy of Artificial Intelligence (BAAI). Guyue Liu is the corresponding author.
\end{acks}

\clearpage{
\bibliographystyle{ACM-Reference-Format}
\bibliography{reference}
}

\clearpage{
\appendix
\onecolumn
\input{appendix-nsdi.tex}
}

\end{document}

%% file: cmds.tex
\newcommand{\ie}{{i.e.,}\xspace}
\newcommand{\eg}{{e.g.,}\xspace}
\newcommand{\etc}{{etc.}\xspace}

\newcommand{\head}[1]{\noindent{\bf #1:}}
\newcommand{\sref}[1]{\S\ref{#1}}

\newcommand{\fix}[1]{\textcolor{black}{#1}}

%% file: macros.tex
\usepackage{hyperref}
\usepackage{enumitem}
\usepackage{multirow}
\usepackage{array}
\usepackage{graphicx}
\usepackage{caption}
\usepackage{subcaption}
\usepackage{tikz}
\usepackage[linesnumbered,ruled,vlined]{algorithm2e}
\usepackage{algorithmicx}
\usepackage[noend]{algpseudocode}
\usepackage{graphics}
\usepackage{epsfig}
\usepackage{bm}
\usepackage{booktabs}
\usepackage{enumitem}
\usepackage{multirow}
\usepackage{array}
\usepackage{graphicx}
\usepackage{caption}
\usepackage{subcaption}
\usepackage{graphics}
\usepackage{epsfig}
\usepackage{booktabs}
\usepackage{pifont}
\usepackage{xcolor, colortbl}
\usepackage{xspace}
\usepackage{listings}
\usepackage{boldline}
\usepackage{multirow}
\usepackage{xspace}
\usepackage[hang,flushmargin]{footmisc}
\usepackage[font=bf]{caption}
\usepackage{soul}
\usepackage{amsmath}
\usepackage{amsthm}
\usepackage{bbm}
\usepackage{tabularx}
\usepackage{minted}
\usepackage{svg}
\usepackage{comment}

\newcommand{\paperid}{xxx}

\newcommand{\sysname}{\textsf{HetCCL}\xspace}

\newcommand{\cluster}{\emph{Cluster}\xspace}

\newcommand{\coll}{\texttt{coll}\xspace}
\newcommand{\sync}{\texttt{sync}\xspace}

\newcommand{\iluvatar}{Vendor~1\xspace}
\newcommand{\cambricon}{Vendor~2\xspace}
\newcommand{\metax}{Vendor~3\xspace}
\newcommand{\accl}{V1CCL\xspace}
\newcommand{\bccl}{V2CCL\xspace}
\newcommand{\cccl}{V3CCL\xspace}

\definecolor{codegreen}{rgb}{0,0.6,0}
\definecolor{codegray}{rgb}{0.5,0.5,0.5}
\definecolor{codepurple}{rgb}{0.58,0,0.82}
\definecolor{backcolour}{rgb}{0.95,0.95,0.92}
\definecolor{brickred}{rgb}{0.8,0.25,0.33}
\definecolor{lightgray}{rgb}{0.83, 0.83, 0.83}

\setlist{nolistsep}
\setlist[itemize]{leftmargin=10pt}
\setlist[enumerate]{leftmargin=13pt,label={(\arabic*)}}

\newcounter{packednmbr}
\newenvironment{packedenumerate}{\begin{list}{\thepackednmbr.}{\usecounter{packednmbr}\setlength{\itemsep}{2pt}\addtolength{\labelwidth}{-6pt}\setlength{\leftmargin}{12pt}\setlength{\listparindent}{\parindent}\setlength{\parsep}{1pt}\setlength{\topsep}{2pt}}}{\end{list}}

\newenvironment{packeditemize}{\begin{list}{$\bullet$}{\setlength{\itemsep}{2pt}\addtolength{\labelwidth}{-6pt}\setlength{\leftmargin}{12pt}\setlength{\listparindent}{\parindent}\setlength{\parsep}{1pt}\setlength{\topsep}{2pt}}}{\end{list}}

\lstdefinestyle{mystyle}{
    backgroundcolor=\color{backcolour},   
    commentstyle=\color{codegreen},
    keywordstyle=\color{magenta},
    numberstyle=\tiny\color{codegray},
    stringstyle=\color{codepurple},
    basicstyle=\ttfamily\footnotesize,
    breakatwhitespace=false,         
    breaklines=true,                 
    captionpos=b,                    
    keepspaces=true,                 
    numbers=left,                    
    numbersep=5pt,                  
    showspaces=false,                
    showstringspaces=false,
    showtabs=false,                  
    tabsize=2,
}
\newtheoremstyle{acmplain}%
  {.3\baselineskip\@plus.2\baselineskip
    \@minus.2\baselineskip}
  {.3\baselineskip\@plus.2\baselineskip
    \@minus.2\baselineskip}
  {\normalfont}
  {0pt}
  {\bfseries}
  {.}
  {.5em}
  {\thmname{#1}\thmnumber{ #2}\thmnote{ (#3)}}

\newtheoremstyle{acmdefinition}%
  {.3\baselineskip\@plus.2\baselineskip
    \@minus.2\baselineskip}
  {.3\baselineskip\@plus.2\baselineskip
    \@minus.2\baselineskip}
  {\normalfont}
  {0pt}
  {\bfseries}
  {.}
  {.5em}
  {\thmname{#1}\thmnumber{ #2}\thmnote{ (#3)}}

%% file: 1-intro.tex
\section{Introduction\label{sec:intro}}

The rapid advancement of Large Language Models (LLMs) has driven two key trends in large-scale training. First, the increasing size of LLMs demands ever-larger GPU clusters. State-of-the-art models have grown from billions to trillions of parameters~\cite{pangu,switch_transformer}, requiring thousands of GPUs to meet their immense computational and memory requirements. Second, enterprises are increasingly adopting diverse GPUs from multiple vendors (\eg NVIDIA~\cite{a100}, AMD~\cite{mi300x}, Huawei~\cite{ascend}). Relying solely on a single vendor to build homogeneous clusters (e.g., NVIDIA's A100 accelerators~\cite{a100}) is becoming increasingly impractical due to factors such as cost efficiency, incremental upgrades, and supply constraints (details in \sref{subsec:motivation-bg}). 

The shift towards multi-vendor heterogeneous clusters presents a promising opportunity to enhance resource utilization and cost-effectiveness for LLM training. Realizing this potential hinges on the efficiency of \textit{collective communication libraries (CCLs)}~\cite{nccl,msccl,oneccl}, which orchestrate essential operations such as AllReduce and AllGather. These operations are fundamental to parallelization strategies~\cite{Megatron-LM,zero} like data parallelism (DP) and tensor parallelism (TP) \etc, ensuring efficient synchronization and data exchange across GPUs. Unfortunately, existing communication libraries are designed for homogeneous environments and struggle to efficiently support heterogeneous clusters.
Existing collective communication libraries broadly fall into two categories:
\begin{packeditemize}
    \item \textit{Device-centric libraries}, such as NCCL from Nvidia~\cite{nccl} and RCCL from AMD~\cite{rccl}, are kernel-based collective algorithm implementations deeply optimized for their respective hardware, utilizing vendor-specific techniques such as NVSHMEM~\cite{nvidia_nvshmem} and GPU-direct RDMA~\cite{nvidia_gdr} for fast intra- and inter-node communication.
    CCLs of this category achieve high efficiency in homogeneous clusters, but are inherently vendor-locked, preventing interoperability with GPUs from other manufacturers. 
    \item \textit{Host-centric libraries}, such as OpenMPI~\cite{openmpi} and Gloo~\cite{gloo}, are designed for traditional HPC workloads and rely on host memory as an intermediate buffer for data transfers.
    While CCLs of this category enable broader hardware compatibility, the frequent host-device memory copies introduce substantial overhead. This inefficiency becomes particularly problematic in large-scale LLM training, where high communication costs can significantly degrade overall performance.
\end{packeditemize}

\input{fig_motivation-heterogeneity}

Neither approach provides an efficient solution for heterogeneous LLM training. Device-centric libraries are restricted to single-vendor environments, while host-centric libraries suffer from excessive communication overhead. As a result, existing solutions force a trade-off between implementation efficiency and cross-vendor compatibility, limiting their practicality in heterogeneous clusters. This reality raises a timely and important question: \textit{How to design a collective communication library that efficiently supports heterogeneous LLM training without compromising performance or efficiency}?

Answering this question requires overcoming challenges at multiple levels, including the data path efficiency for device-to-device communication, implementation portability of data reduction across platforms, and the abstraction for heterogeneous cluster topologies and collective algorithms.
In this paper, we propose \sysname, a novel framework that enables the modeling and optimization of collective communications in mixed-vendor accelerator card clusters.
\sysname is built on the following key ideas:

\begin{packeditemize}
    \item \emph{Cross-vendor Device Data Transport:} Existing data exchange in collective communications either copies data to a CPU bounce buffer (CPU-forwarding approach~\cite{openmpi,gloo}) or utilizes vendor-specific transports on device memory buffers, such as NVLink and Infinity Fabric for intra-server interconnect, and GPU-Direct RDMA (GDR~\cite{hamidouche2015exploiting,potluri2013efficient}) for inter-server RDMA transport.
    The CPU-forwarding approach suffers from significant host-device memory copy overhead, and the vendor-specific transports embedded in collective communication kernels are not portable across platforms.
    We decouple the control logic and data path of the device buffer RDMA, using a host-centric, kernel-free control logic to ensure vendor compatibility, while keeping the data path completely on-device, thereby eliminating the host-memory copying overhead associated with CPU-based approaches.

    \item \emph{Vendor-independent Reduction:}
    Apart from data movements, combining collectives (\ie AllReduce, ReduceScatter, and Reduce) also performs data computation.
    Given the programmability discrepancies across different vendor hardware in heterogeneous clusters, it is challenging to implement data reduction compatible with a wide range of vendors.
    Although existing work has proposed offloading reduction to the CPU~\cite{an2024fire}, this approach is against the design principle of keeping the data path on-device.
    Our design utilizes the device computation resource for faster data reduction and lower data copy overhead.
    We propose a cross-vendor data exchange pattern that stores local and received data on separate ranks and uniformly implements reduction utilizing the reduce interface provided by existing homogeneous collective libraries.
    
    \item \emph{Hierarchical Algorithm Design:}
    In mixed-vendor clusters, the key challenge lies in the trade-off between algorithm flexibility and implementation efficiency.
    Using peer-to-peer (P2P) data transfer as the core primitive for collective communications offers high algorithm flexibility (\eg existing works on automatic collective algorithm optimization~\cite{zhao2024forestcoll,shah2023taccl,kim2024tccl,liu2024rethinking}).
    On the other hand, homogeneous device-centric collective implementations provide higher implementation efficiency.
    We propose a fused approach based on our hierarchical topology abstraction that divides the heterogeneous cluster into multiple homogeneous clusters.
    At the homogeneous cluster level, we propose an algorithm primitive abstraction including both intra- and inter-cluster operations, leveraging the flexibility of cluster-level P2P primitives and the efficiency of device-centric homogeneous collective primitives.

    \item \emph{Pipelined Collective Algorithm Execution:}
    Sequentially executing the algorithm primitives would lead to bandwidth underutilization, as the inter-cluster links remain idle while waiting for intra-cluster operations to complete.
    On top of the primitive abstraction and collective algorithm design, we build a pipelined execution workflow to overlap intra- and inter-cluster primitives, maximizing bandwidth utilization across the heterogeneous cluster.
\end{packeditemize}

We implement \sysname with $30$k LOC, supporting 8 different vendors, and incorporate \sysname into PyTorch via the customized backend. We evaluated it on a heterogeneous cluster containing hardware from 4 vendors with P2P SendRecv, collective communication benchmarks, as well as end-to-end performance, including training with Llama3-3B/8B models and serving with Qwen2-7B model. To the best of our knowledge, this is the first heterogeneous collective communication library that supports full device-buffer data transfer and MPI-style collective operations. \sysname is publicly available. We omit the link to comply with anonymity requirements.

Our evaluation results show that \sysname achieves up to $91.4\%$ hardware bandwidth of the slowest vendor in heterogeneous SendRecv and $97\%$ the bandwidth of homogeneous collectives in heterogeneous collective communications.
In end-to-end training, \sysname accelerates the per-step-time by $9.1\%$ and $16.9\%$ for Llama3-3B and Llama3-8B models, where the computation dominates communication. We expect even greater benefits from \sysname in more communication-intensive scenarios, such as large-scale model training~\cite{llama3herdmodels,deepseekv3}.

\noindent \emph{Ethics:} This work does not raise any ethical issues.

%% file: fig_motivation-heterogeneity.tex
\begin{figure*}[!tbp]
    \centering
    \includegraphics[width=\textwidth]{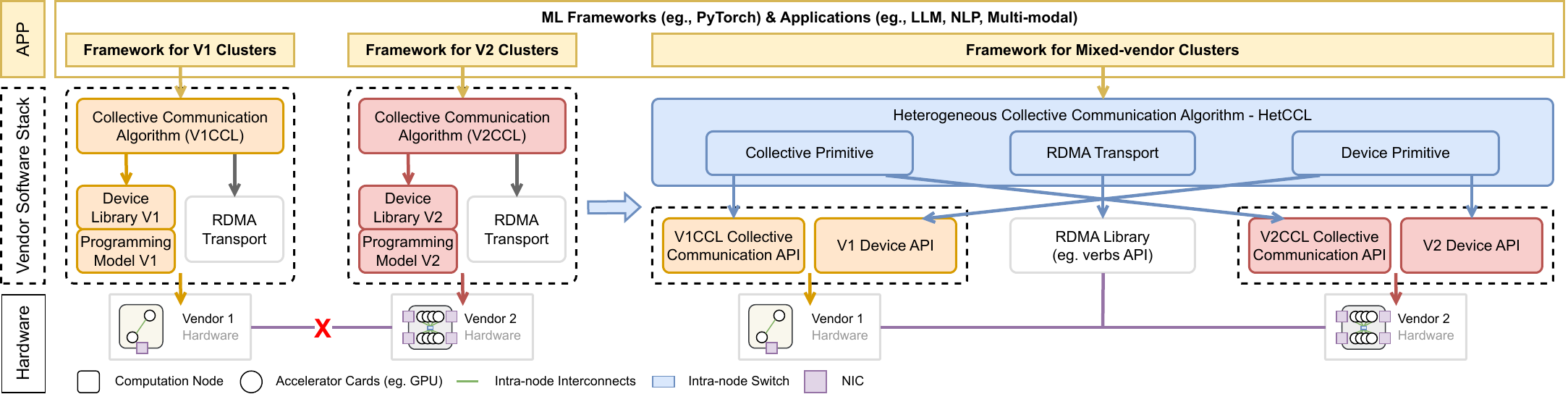}
    \caption{
    \small Existing and our proposed architecture for clusters with hardware from multiple vendors, dealing with the hardware heterogeneity, differences in the programming models, and varying visibility and control over the underlying hardware.
    }
    \label{fig:motivation-heterogeneity}
\end{figure*}

%% file: 2-motivation.tex
\input{fig_motivation-data-path}

\section{Background and Motivation\label{sec:motivation}}
In \sref{subsec:motivation-bg}, we introduce the necessity of heterogeneous collective communication and the requirements for an ideal solution.
In \sref{subsec:motivation-challenge-1}$\sim$\sref{subsec:motivation-challenge-3}, we analyze the limitations of existing approaches and the challenges in meeting the requirements.

\subsection{Heterogeneous Collective Communication\label{subsec:motivation-bg}}

\head{LLM training in heterogeneous accelerator clusters}
Enterprises are increasingly adopting heterogeneous hardware~\cite{hong2022dfx,kachris2025survey,chen2024understanding,huang2024new} for large language model (LLM) training for practical reasons.
First, the diversity of hardware vendors and architectures has expanded significantly over the years.
Beyond major vendors like NVIDIA and AMD, other vendors (\eg Huawei~\cite{ascend}, Graphcore~\cite{graphcore}) offer enterprises broader options considering availability and task-specific optimizations. Table~\ref{tab:motivation-vendors} presents representative hardware settings, with vendor names anonymized due to confidentiality considerations.
Second, the growing computational demands of state-of-the-art LLMs necessitate scalable infrastructure, which is rarely deployed in a uniform, single-generation manner.
Instead, clusters often evolve incrementally, reflecting the natural life cycle of data centers where multi-generational hardware must coexist and interoperate effectively.
Moreover, in commercial server rental and cloud provisioning scenarios, it is often difficult for users to acquire a large number of homogeneous server instances concurrently due to dynamic availability and fragmented resource supply, which can constrain the scale of LLM workloads or incur substantial waiting time.
As a result, integrating heterogeneous and fragmented available compute resources for LLM training and inference becomes a practical approach to improving overall resource utilization in cloud environments.
Furthermore, enterprises leverage heterogeneous cluster setups to avoid being limited by a single vendor’s hardware production capacity, ensuring procurement flexibility while maximizing the utility of legacy hardware.
This blend of scalability, adaptability, and optimization underscores the increasing reliance on heterogeneous clusters in modern LLM workloads.

\input{tab_motivation-vendors}

\head{Heterogeneous collective communications}
Collective communication libraries (CCLs) play a pivotal role in maximizing the efficiency of heterogeneous hardware by coordinating essential operations such as AllReduce and AllGather.
\fix{Some collective operations only involve data movements, which are commonly referred to as non-combining collectives~\cite{cai2021synthesizing}.
On the other hand, AllReduce, ReduceScatter, and Reduce operations collect multiple versions of data and produce a reduced output, which are referred to as combining collectives.}
As listed in Table~\ref{tab:motivation-coll}, these operations serve as the backbone of widely adopted parallel training strategies, \eg data parallelism (DP), pipeline parallelism (PP), and tensor parallelism (TP)~\cite{Megatron-LM,zero}, facilitating efficient synchronization and data exchange across multiple GPUs.

\input{tab_motivation-coll}

Unfortunately, existing communication libraries are designed for homogeneous environments, as shown in Figure~\ref{fig:motivation-heterogeneity}~(left). 
\fix{Applications such as ML frameworks adapt to each vendor CCL as separate backends to achieve portability across homogeneous clusters of different vendors, but cannot run across a heterogeneous cluster due to incompatible CCL backends.}
More specifically, vendor-specific CCLs implement collective algorithms (the scheduling of data movements, reductions, and synchronizations) with tailored device code, using specialized low-level device libraries and programming models (\eg CUDA and ROCm) for different types of hardware.
This lack of interoperability makes it difficult for upper-layer frameworks to adapt to new hardware vendors or apply optimizations across mixed-vendor clusters.

An ideal solution for heterogeneous collective communication should meet the following requirements: 
1) \textit{portability} across heterogeneous hardware,
2) \textit{seamless integration} with downstream applications,
3) \textit{minimum software adaptation} required from each vendor,
4) \textit{high performance} in collective communication.
Fulfilling these requirements calls for an alternative architecture, as shown in Figure~\ref{fig:motivation-heterogeneity} (right), with a collective communication library that wraps the underlying heterogeneity across hardware vendors and exposes high-performance collective interfaces to applications.
We observe that the hardware heterogeneity across vendors includes their accelerator card model and intra-server high-bandwidth interconnects, but they share the same scale-out network architecture, namely the inter-server RDMA (Remote Direct Memory Access) transport.
Therefore, we choose the common RDMA APIs (\ie verbs) as the bridge for cross-vendor device data transport, with a minimal set of unified device APIs universally provided by various vendors.
Building a collective communication library with our proposed architecture faces several challenges, spanning from the data transfer primitive to the design of the collective algorithms.

\subsection{Heterogeneous Data Transfer Challenge~\label{subsec:motivation-challenge-1}}

\head{Challenge 1} \emph{How to perform efficient and kernel-free device data transfer between heterogeneous peers.}

We group common collective communication libraries into device-centric and host-centric approaches based on their different inter-node data transfer mechanisms~\footnote{Accelerator cards are provided and purchased as integrated servers, so hardware within the same node is homogeneous, and heterogeneous peers only exist in inter-node data transfers}.
Table~\ref{tab:motivation-summary} summarizes the characteristics of the two approaches.

\input{tab_motivation-summary}

\head{Device-centic approaches}
Vendor-developed CCLs, such as NCCL~\cite{nccl} for NVIDIA and RCCL~\cite{rccl} for AMD, are meticulously optimized for homogeneous environments.
These libraries leverage in-depth knowledge of their respective hardware architectures and programming models to achieve high performance.
Their inter-node data transfer typically follows Figure~\ref{fig:motivation-data-path}~(a), where the data movements are device-driven and directly in device memory buffers via GPUDirect RDMA~\cite{hamidouche2015exploiting,potluri2013efficient}, ROCm~\cite{rocm}, \etc.
This mechanism avoids cost-inefficient operations such as host-device memory copies.

The limitation is the strong dependency on vendor-specific technologies, which limits their cross-vendor compatibility.
Their only support of hardware heterogeneity is across different hardware generations from a single vendor.

\head{Host-centric approaches}
Another line of collective communication implementations, such as OpenMPI~\cite{chen2023mpi} and Gloo~\cite{gloo}, originally focuses on distributed CPU tasks, such as traditional HPC applications.
While some have been extended to support GPU transport via third-party integrations, their core architecture remains host-centric.
These libraries can support cross-platform device data exchange, as shown by Figure~\ref{fig:motivation-data-path}~(b), where the device buffer is copied to a host bounce buffer and forwarded to the remote node's host buffer via TCP or CPU RDMA, then copied to the peer device buffer.

This mechanism falls short in its data-path efficiency, as data must frequently move between the device and the host, incurring heavy memory-copy overhead and PCIe bottleneck.
For modern workloads such as LLM training that demand high-speed collective operations, this can become a significant bottleneck.

\head{Data path efficiency comparison}
\fix{Figure~\ref{fig:motivation-perf-comparison} shows the memory copying time in different data path implementations, profiled during 2GB SendRecv operations of NVIDIA (\texttt{nv}) and Vendor 1 (\texttt{v1}) hardware.}
For mechanism~(b) in Figure~\ref{fig:motivation-data-path}, a device-to-host memory copy (\texttt{d2h}) on the sender and a host-to-device memory copy (\texttt{h2d}) on the receiver takes up more than $3.8\times$ time on average than two device-to-device (\texttt{d2d}) memory copies.
On the other hand, a fully device-driven method would require a unified and efficient programming model with device RDMA support (\eg CUDA and GDR for NVIDIA hardware) that is compatible with every vendor, which lacks established standards for varying hardware architectures and abilities.
Therefore, keeping an on-device data path for heterogeneous data transfer is challenging, as it requires balancing compatibility and performance between the two existing inter-node data transfer mechanisms.

\input{fig_motivation-perf-comparison}

\subsection{Portable Reduction Challenge\label{subsec:motivation-challenge-2}}

\noindent\textbf{Challenge 2:} \textit{Implement vendor-agnostic data reduction}.

In \sref{subsec:motivation-challenge-1}, we focus on the efficiency of data transfers across heterogeneous peers, which is sufficient to construct non-combining collective operations (Table~\ref{tab:motivation-coll}), but combining collectives such as AllReduce and ReduceScatter are also important components in downstream tasks, which additionally demand the ability to reduce data (\eg compute sum, min, and max).
This introduces new challenges.
\fix{Existing reduction implementations rely on the portability of their underlying programming model to scale to more hardware types, which lacks a universal solution.}
\fix{Writing separate device code for each distinct device model requires vigorous and continuous effort, and does not apply to vendors that do not provide an open programming platform.}
A recent work, HFReduce~\cite{an2024fire}, offloads data reduction to the CPU, potentially making it portable across vendors.
But this is incompatible with the goal of maintaining an on-device data path for heterogeneous collective communication.
The challenge lies in reducing device data in a uniform way across various vendors in heterogeneous combining collectives.

\subsection{Collective Algorithm Efficiency Challenge\label{subsec:motivation-challenge-3}}
\noindent\textbf{Challenge 3:} \textit{Efficiency trade-off between collective algorithm flexibility and implementation efficiency}.

The collective algorithm determines the theoretical communication performance.
Industrial solutions such as NCCL~\cite{nccl} implement limited pre-defined algorithms, such as the ring or tree algorithm.
Downstream applications such as LLM training frameworks use these libraries via the exposed APIs in a black-box manner.
However, pre-defined algorithms do not guarantee optimal latency or bandwidth for arbitrary topologies, especially in heterogeneous settings.
Recent works~\cite{cai2021synthesizing,shah2023taccl,liu2024rethinking} search for optimal customized algorithms under their communication cost models, using P2P data transfer and device data reduction as algorithm primitives.
Collective operations, namely scheduling of these primitives, run in a high-performance execution backend~\cite{msccl,cai2021synthesizing} implemented for homogeneous NVIDIA or AMD hardware only.
For heterogeneous clusters, completely relying on kernel-free P2P data transfer introduces higher overhead across homogeneous peers than existing kernel implementations.

In a word, vendor-provided CCL implementations fall short in cross-vendor portability and algorithm flexibility.
Modeling-based collective optimization approaches provide a more flexible algorithm design space, but when executed with heterogeneous P2P transport primitives, yield higher implementation overhead.
We need to balance the trade-off between these two approaches in our heterogeneous collective algorithm
design, achieving both the high implementation efficiency of device kernels and the flexibility of P2P data transfers.

%% file: fig_motivation-data-path.tex
\begin{figure*}[!tbp]
    \centering
    \includegraphics[width=\textwidth]{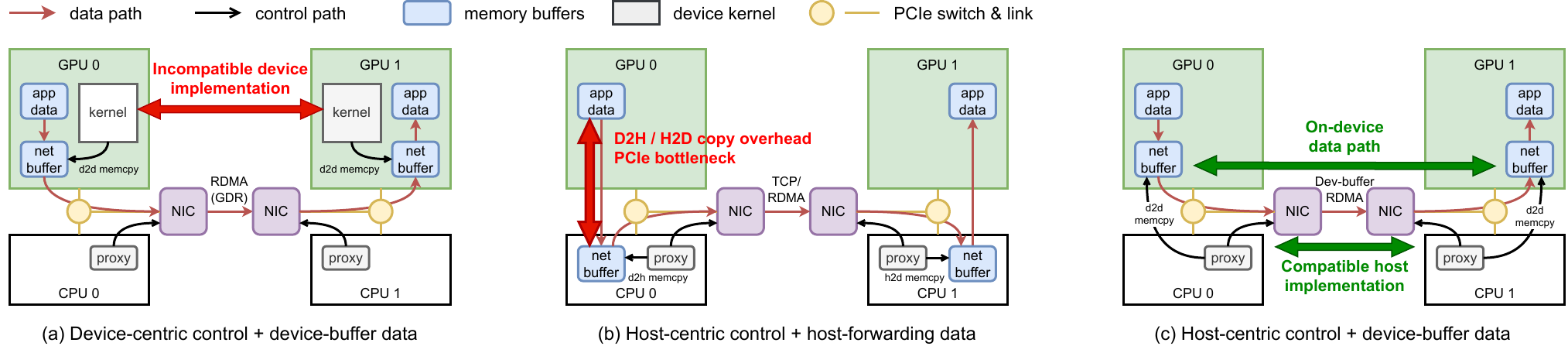}
    \caption{\small Inter-node device data transfer mechanism comparison.}
    \label{fig:motivation-data-path}

\end{figure*}

%% file: tab_motivation-vendors.tex
\begin{table}[!thbp]
\centering
\resizebox{\columnwidth}{!}{
\begin{tabular}{c|c|c|c|c}
    \hline
    \textbf{Hardware} & \textbf{\#dev} & \textbf{TFLOPS (FP32)} & \textbf{Scale-up Network} & \textbf{Scale-out Network} \\
    \hline\hline
    \textbf{NVIDIA A100} & 8 & 156 TFLOPS & 4.8 TB/s NVLink & 8x200G IB \\ \hline
    \textbf{AMD MI300x} & 8 & 163.4 TFLOPS & Infinity Fabric & configurable IB \\ \hline
    \textbf{Intel Gaudi 3} & 8 & 1835 TFLOPS (FP16) & 4.2 TB/s RoCE & 3x300G RoCE \\ \hline
    \hline
    \textbf{\iluvatar} & 16 & 32 TFLOPS & 192GB/s & 100G IB \\ \hline
    \textbf{\cambricon} & 8 & 512 TOPS (INT8) & 8x100GB/s & 2x200G IB \\ \hline
    \textbf{\metax} & 8 & 200TFLOPS & 240GB/s & 2x400G IB \\ \hline
\end{tabular}}
\caption{\small Summary of various accelerator card vendors and servers. Today's major hardware vendors include NVIDIA, AMD, and Intel.
}
\label{tab:motivation-vendors}
\end{table}

%% file: tab_motivation-coll.tex
\begin{table}[!thbp]
\centering
\resizebox{\columnwidth}{!}{
\begin{tabular}{c|c|c}
    \hline
    \textbf{Type} & \textbf{Collective Operation} & \textbf{Parallel Strategy} \\ \hline
    \hline
    Non-combining & SendRecv (P2P) & Pipeline Parallelism (PP) \\ \hline
    Non-combining & AllGather      & Tensor Parallelism (TP) \\ \hline
    Combining     & AllReduce      & Data and Tensor Parallelism (DP, TP) \\ \hline
    Combining     & ReduceScatter  & Tensor Parallelism (TP) \\ \hline
\end{tabular}}
\caption{\small Summary of collective communications used in common parallel strategies (DP, TP and PP).
}
\label{tab:motivation-coll}
\end{table}

%% file: tab_motivation-summary.tex
\begin{table}[!thbp]
\centering
\resizebox{\columnwidth}{!}{
\begin{tabular}{c|c|c|c|c}
    \hline
    \textbf{CCL} & \textbf{Performance} & \textbf{Compatibility} & \textbf{Data Path} & \textbf{Control Path} \\ \hline
    \hline
    \textbf{NCCL} & High & Single vendor & GPU-buffer RDMA & GPU kernel \\ \hline
    \textbf{RCCL} & High & Single vendor & GPU-buffer RDMA & GPU kernel \\ \hline
    \hline
    \textbf{OpenMPI} & Low & Mixed vendor & CPU forwarding & CPU control \\ \hline
    \textbf{Gloo} & Low & Mixed vendor & CPU forwarding & CPU control \\ \hline
    \hline
    \textbf{\sysname} & \textit{Relatively High} & \textit{Mixed vendor} & \textit{GPU-buffer RDMA} & \textit{CPU control} \\ \hline
\end{tabular}}
\caption{\small Summary of existing frameworks on their vendor compatibility, performance, and the data and control path implementation for inter-node data transfer.
}
\label{tab:motivation-summary}
\end{table}

%% file: fig_motivation-perf-comparison.tex
\begin{figure}
    \centering
    \includegraphics[width=.9\columnwidth]{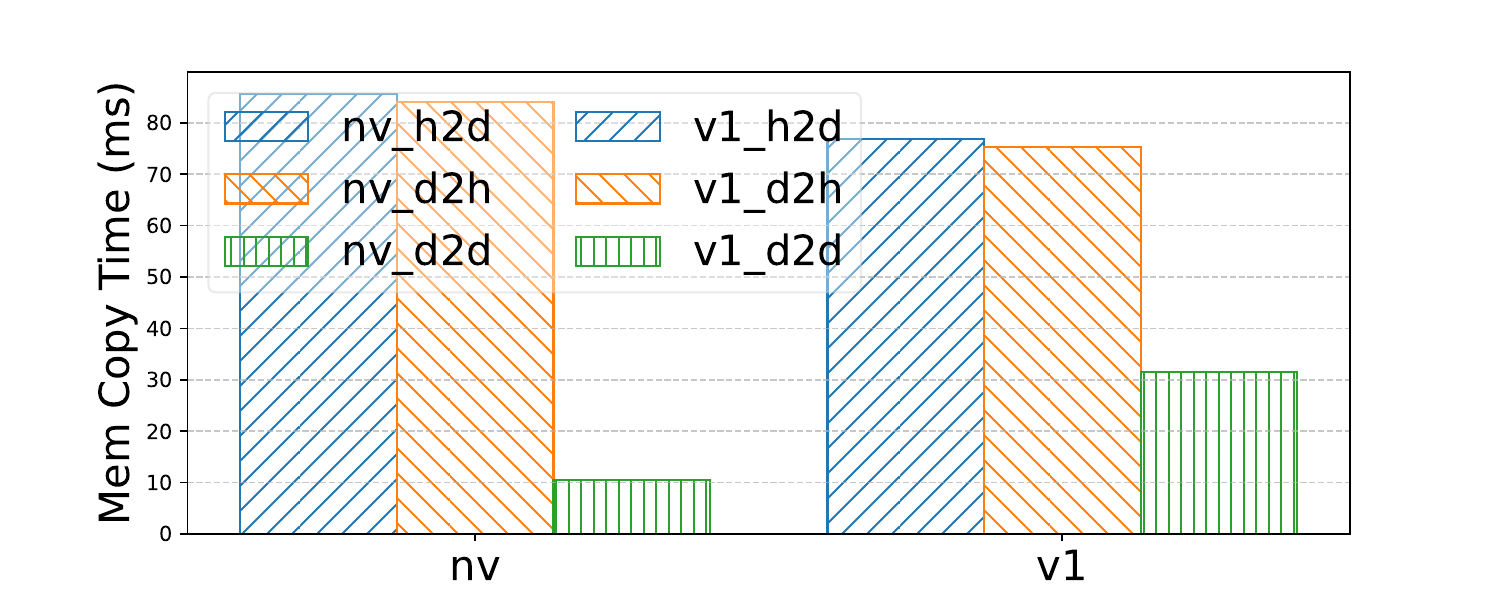}
    \caption{\small Data path overhead of different mechanisms.}
    \label{fig:motivation-perf-comparison}
\end{figure}

%% file: 3-overview.tex
\section{Design Overview\label{sec:overview}}

In this section, we present the key insights to address the above challenges for heterogeneous collective communication, followed by an overview of the \sysname architecture.

\subsection{Key Ideas\label{subsec:overview-insight}}

\head{Idea 1. Decouple data and control paths for \fix{P2P transfer of device data}}
We separate the RDMA data and control paths to eliminate the host-device data copy overhead for heterogeneous peers.
\sysname adopts a novel host-driven device-buffer RDMA mechanism, where the control logic (\eg memory region management, RDMA operations, connection management, event handling, \etc) is scheduled on the host side for maximum hardware and programming compatibility, and the data path remains on the device to avoid host-device data movement and the potential PCIe bottleneck (Figure~\ref{fig:motivation-data-path} (c)).

\head{Idea 2. Vendor-independent reduction using native combining collectives}
Vendor-provided CCLs provide built-in kernel implementations of data reduction for combining collectives.
By leveraging the multiple ranks involved in a collective operation, we insert an intermediate receiving rank and align the offsets for P2P data transfers to perform the data computation using built-in reduction implementations in existing CCLs.

\head{Idea 3. Hierarchical collective algorithm design based on vendor hardware groups}
Existing vendor CCLs are more efficient for homogeneous device subgroups, whereas the P2P-level data transfer primitive offers greater flexibility and better algorithmic optimality.
We propose a hierarchical topology abstraction and a corresponding collective algorithm design, applying high-efficiency device kernels within homogeneous device subgroups in the heterogeneous cluster, while leveraging the flexibility of P2P data transfers to meet the remaining communication requirements.

\input{fig_overview}

\subsection{\sysname Architecture Design\label{subsec:overview-architecture}}

Figure~\ref{fig:overview} shows the design overview of \sysname, including three key components and a wrapper module:
1) a device-level primitive abstraction for efficient heterogeneous P2P transport,
2) a hierarchical topology abstraction and corresponding cluster-level primitives,
3) the heterogeneous collective algorithm with a pipelined execution workflow,
and 4) a light-weight vendor API wrapper module for leveraging vendor CCL optimizations. 

\begin{packedenumerate}
\item[\textbf{\ding{172}}] \textbf{\fix{P2P transport for device data} (\sref{subsec:design-1}):}
We summarize a list of node-level primitives (Table~\ref{tab:rdma-primitives}) to enable host-driven device-buffer heterogeneous P2P transport.
The primitives include device memory operations, control operations (\eg stream management), host-device coordination, and RDMA resource management.

\item[\textbf{\ding{173}}] \textbf{Cluster-to-cluster primitives (\sref{subsec:design-2}):}
Our hierarchical topology abstraction breaks the heterogeneous cluster into homogeneous subgroups, enabling homogeneous collective semantics and group data transfer semantics via the cluster-level primitives (Table~\ref{tab:collectives-breakdown}).
We ensure the efficiency of \textit{cross-vendor communication} primitives via multi-channel load-balanced P2P data transfer.
For combining collectives, we design a \textit{data transfer-reduce} primitive that uses the native homogeneous combining collective implementation to produce the reduction result for the global collective operation.

\item[\textbf{\ding{174}}] \textbf{Collective algorithm and pipelined execution (\sref{subsec:design-3}):}
We propose a collective operation breakdown (Algorithm~\ref{algo:c2c-breakdown}) aligned with the hierarchical topology, where cluster-level primitives collectively implement the semantics of heterogeneous collective operations.
The collective algorithm efficiency is guaranteed by optimal cross-vendor data transfer volume and pipelined execution workflow.
\end{packedenumerate}

%% file: fig_overview.tex
\begin{figure}[tbp]
    \centering
    \includegraphics[width=\columnwidth]{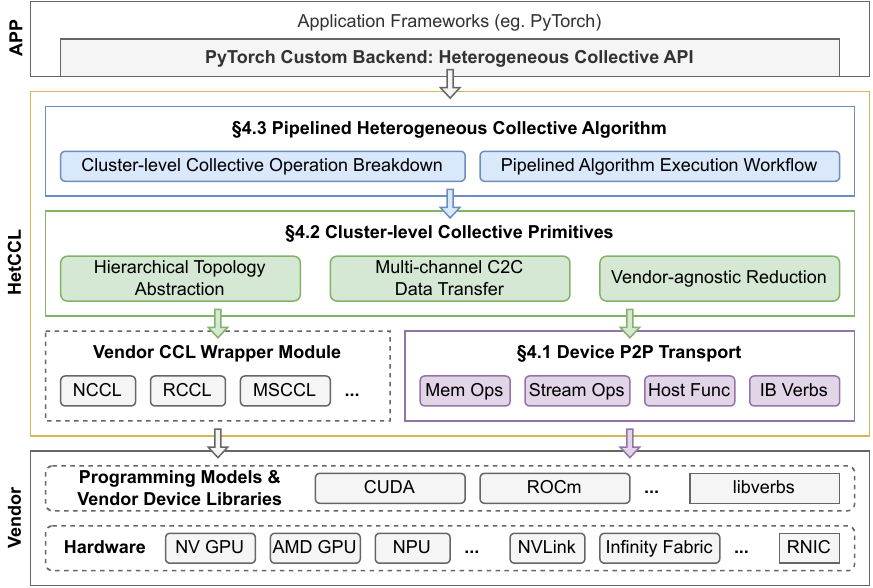}
    \caption{
    \small Design Overview
    }
    \label{fig:overview}
\end{figure}

%% file: 4-design.tex
\section{\sysname Design\label{sec:design}}

In \sref{subsec:design-1}, we introduce our heterogeneous device-buffer P2P transport design.
\sref{subsec:design-2} presents the cluster-level primitives and explains how \sysname addresses the compatibility issue encountered by combining collectives.
\sref{subsec:design-3} introduces the hierarchical collective algorithm and its pipelined execution.

\subsection{\fix{Device Buffer P2P Transport} \label{subsec:design-1}}

\input{fig_design-gdr}

The basic building block of heterogeneous collective communication is the underlying heterogeneous P2P transport.
\sysname enables \textit{device buffer data transfer} across heterogeneous peers via a novel host-device-coordinated RDMA mechanism.
As summarized in \sref{subsec:motivation-challenge-1}, the data transfer across heterogeneous devices in existing frameworks copies device buffers to host buffers and uses host transport (RDMA or TCP) to transfer data across heterogeneous nodes, which suffers from significant memory copying overhead and PCIe bottleneck.
\sysname proposes mechanism~(c) of Figure~\ref{fig:motivation-data-path} to follow the data path of mechanism~(a) while keeping the cross-vendor portability of mechanism~(b).
More specifically, mechanism~(c) is built with the following components:

\head{Host-driven control logic}
Based on the common host functionalities for managing device memory and RDMA operations across various vendors, instead of the device-centric data transfer logic (\eg NCCL~\cite{nccl}, RCCL~\cite{rccl}), we offload the RDMA control logic to a host proxy, guaranteeing high vendor compatibility.
The host proxy functionalities include initializing connections, resource management (\eg managing memory region (MR), completion queue (CQ), protection domain (PD), RDMA buffers, \etc), memory operations (\eg \texttt{malloc} and \texttt{memcpy}), and RDMA operations (\eg posting Work Requests (WR) and polling from the CQ).

When the send host proxy handles a data transfer job, it first calls \texttt{d2dMemcpy} to move the target data to an available RDMA send buffer on the device.
The proxy queries the stream state for the completion of \texttt{d2dMemcpy}, after which the proxy enqueues the send WR to the send queue of the QP to the target receiver.
On the receiver side, the host proxy receives data by polling the RDMA receive queue, after which the receiver buffer is copied to the target device buffer.

\head{On-device data path}
This mechanism keeps the RDMA data path fully on-device: 1) the sender copies data from the device memory to the RDMA send buffer, 2) the RNIC transfers the RDMA buffer to the remote peer through the RNIC, and 3) the receiver copies the receive buffer to the device memory.
Compared with the CPU-forwarding mechanism, host-device memory copies are replaced with device-to-device memory copies (similar to existing device-centric solutions), eliminating the most significant data-path overhead.
\sysname further pipelines the above control logic to overlap the memory-copying and RDMA transfer time and to reuse a pre-allocated RDMA buffer pool.
A data transfer job is sliced into fixed-sized data chunks, and the \texttt{memcpy} and RDMA operations are launched asynchronously (\eg in Figure~\ref{fig:design-gdr}, Data-1 does not require the completion of Data-0 to start its \texttt{memcpy}).

\head{Host-device coordination}
In addition to satisfying the data movement semantics, the P2P primitive also guarantees operational integrity and maintains the same device-side dependency as a device operation.
For instance, jobs submitted to a device stream should execute sequentially.
However, since \sysname offloads the control logic to the CPU, there is no control instance on the device execution pipeline (\eg CUDA stream) to ensure that future jobs wait for previous RDMA operations to complete.
Figure~\ref{fig:design-gdr} shows the interaction between the host and the device side to maintain the correct dependency while pipelining the memory copy and RDMA transfer in a P2P operation.
The proxy simultaneously calls the \texttt{launchHostFunc} API to serve as a placeholder on the device communication stream, marking events for synchronization and maintaining dependencies.
The proxy polls the CQ for completion of the RDMA WR, after which it can mark the send task as complete and safely release the related RDMA resource for future operations.

\head{Takeaway}
By decoupling the control and data path in RDMA transport into separate implementations on the host and device, \sysname removes the host-device memory-copying overhead and the PCIe bottleneck without device-dependent logic.
\sysname ensures the correct dependency for non-blocking execution, thereby achieving high performance and compatibility across various hardware vendors.

\subsection{Hierarchical Topology Abstraction and Cluster-to-cluster Primitives\label{subsec:design-2}}

\sysname comprises a hierarchical topology abstraction (\sref{subsubsec:design-2-topology}) and cluster-level primitives (\sref{subsubsec:design-2-c2c-prim}), combining heterogeneous P2P RDMA with existing homogeneous communication to achieve a balance between algorithm flexibility and implementation efficiency.

\subsubsection{Heterogeneous topology abstraction\label{subsubsec:design-2-topology}}

\input{fig_design-topo-abst}

\sysname's hierarchical topology abstraction groups the devices into homogeneous sub-clusters, connected by cross-cluster RDMA channels.
Figure~\ref{fig:design-topo-abst} shows the hierarchical abstraction for heterogeneous clusters and the corresponding software structures.
During the global communicator (denoted by $Comm_H$) initialization, the cluster uses a CPU-based bootstrapping network to gather the rank information globally (usually, each accelerator card is initialized as a rank).
Then, \sysname groups ranks (devices) by vendors, which are also the maximal device subsets that can execute kernel-based collective operations.
For instance, Vendor~2 includes 4 nodes, each with 4 accelerator cards and 2 NICs.
A vendor device group can further be divided into disjoint sub-clusters, each initialized with a homogeneous communicator.
For simplicity, in this paper, we use the term \cluster to refer to the sub-clusters, in which vendor-provided libraries can perform homogeneous collectives.
As the intra-node topologies of different vendors are not identical, for instance, the number of devices per node and NICs per node may vary, we define the ranks that have the minimum NUMA distance to an RDMA NIC as \textit{border ranks} and others as \textit{internal ranks}.
Additionally, we create an internal border communicator ($Comm_B$) for the \textit{border ranks} of each \cluster.
For the example in Figure~\ref{fig:design-topo-abst}, Vendor~2 is further divided into two symmetric {\cluster}s, each with a border communicator consisting of 4 ranks.

Every collective operation can be viewed as a communication requirement of data chunks~\cite{cai2021synthesizing}, decomposing a collective communication into intra- and inter-cluster data transfer requirements.
As demonstrated by the red arrows in Figure~\ref{fig:design-topo-abst}, each global collective operation can break down into 3 steps, namely \textbf{a)} \texttt{start} intra-cluster (homogeneous) operations, \textbf{b)} cluster-to-cluster (C2C) data transfers, and \textbf{c)} \texttt{end} intra-cluster (homogeneous) operations.
We define three cluster-level primitives to perform at each step in the next section.

\subsubsection{Cluster-level Primitives\label{subsubsec:design-2-c2c-prim}}

\input{tab_design-c2c-primitives}
\input{fig_design-c2c-algo-prim}

Combining the vendor-provided homogeneous collectives and the flexibility of heterogeneous P2P data transfer in \sref{subsec:design-1}, \sysname can cover the data movements required by global collective communications in a heterogeneous cluster.
Table~\ref{tab:c2c-primitives} defines our cluster-level primitives:
\begin{packeditemize}
\item \emph{homColl} performs a homogeneous collective communication with customized send-buffer and receive-buffer offsets and buffer lengths (\sref{sec:impl}).

\item \emph{c2cCpy} transfers required data from the border ranks of the source cluster to the border ranks of the destination cluster. The data is divided proportionally to the NIC bandwidth and scattered among the receiving border ranks, guaranteeing load balance and optimal transfer volume.

\item \emph{c2cRed} additionally performs a combining collective among the \textit{border ranks} of the destination cluster, achieving a vendor-agnostic implementation of reduction.
\end{packeditemize}
The key challenge to decomposing a collective operation lies in 1) ensuring optimal cross-cluster data transfer volume during step \textbf{b)} data exchange and 2) implementing vendor-agnostic reduce operations for combining collectives.

\head{\texttt{c2cCpy} primitive\label{subsubsec:design-2-c2ccpy}}
\sysname ensures optimal cross-cluster data transfer volume and maximizes the bandwidth utilization of multiple cross-cluster channels and border ranks.
\sysname adopts a cluster-level ring algorithm (only exchanging data with the previous and next cluster) to minimize the total cross-cluster data transfer volume.
During each \texttt{c2cCpy} primitive call, only one data copy is transferred across clusters (\ie the set of received data is distributed across the border ranks of the receiving cluster), while dividing data transfer workload proportionally to the cross-cluster channel bandwidth (\eg RNIC bandwidth of border ranks).
Figure~\ref{fig:design-c2c-cpy} shows an example of \texttt{c2cCpy} primitive, where data $0 \sim 3$ from \cluster~0 and $4 \sim 7$ from \cluster~1 are exchanged across their border ranks.

\head{\texttt{c2cRed} primitive\label{subsubsec:design-2-c2cred}}
For combining collectives, a key challenge to reducing data on the receiver side is cross-platform compatibility.
Especially for emerging vendors, providing a collective communication library implementation on their hardware is significantly easier—and more common in practice—than developing a comprehensive programming platform (\eg CUDA).
This means that it is not always convenient to handcraft a reduction kernel for each type of device in the cluster.
Although moving the reduction to the CPU would be a feasible solution, this violates our key insight of keeping the data path on-device.
Figure~\ref{fig:design-c2c-red} illustrates a running example of \texttt{c2cRed}.
Suppose that some partially reduced data are distributed on multiple ranks in \cluster~0, leaving other offsets currently unoccupied.
The data transfer is similar to \texttt{c2cCpy}, except that the data received from \cluster~1 are routed to ranks where the corresponding data offset is available, then reduced to the target rank and offset by performing a \texttt{Reduce} in the border communicator\footnote{This approach cannot work with single-rank vendor clusters, which we have never encountered in production clusters. In case this happens, we also implement a fallback solution that offloads data reduction to the CPU.}.

\input{algo}

\subsection{\fix{Hierarchical Collective Breakdown Algorithm and Pipelined Execution\label{subsec:design-3}}}

\subsubsection{Cluster-level collective algorithm and communication primitives\label{subsubsec:design-3-c2c-algo}}
In this section, we elaborate on the \emph{C2C collective algorithm} design of \sysname, which performs a global collective operation with \emph{cluster-level} primitives.

Algorithm~\ref{algo:c2c-breakdown} shows the 3-step hierarchical breakdown of collective operations using these primitives.
Line $3 \sim 4$ perform a group of intra-cluster homogeneous collective operations ($startColl$) by calling collective APIs provided in vendor CCLs.
Line $5 \sim 12$ performs \emph{C2C} data transfers, such that each cluster's border ranks will possess all necessary data from other clusters to complete the global collective operation.
Line $13 \sim 14$ wraps up the collective communication with another group of intra-cluster collective operations ($endColl$), generating the final output value to the internal ranks in the cluster.
According to the global communication pattern and the cross-cluster data transfer requirement for each type of collective operation, we summarize the intra-cluster start and end collectives in Table~\ref{tab:collectives-breakdown}, together with two concrete examples of our algorithm (Figure~\ref{fig:design-c2c-ag} and Figure~\ref{fig:design-c2c-ar}), in Appendix~\ref{append:coll-algo}.

\input{fig_design-pipeline}

\subsubsection{Pipelined Execution Workflow\label{subsubsec:design-3-pipeline}}

At a high level, Algorithm~\ref{algo:c2c-breakdown} represents the construction logic of collective operation with our primitives, but executing the loops sequentially results in low hardware utilization.
Take \texttt{AllGather} as an example, we may observe in a cluster that receiving data from $Cluster_i$ and the intra-cluster broadcasting of data already received from $Cluster_{i-1}$ has no data dependency or bandwidth contention.
Furthermore, it is rather common in modern data centers to have every rank in a cluster be a border rank, in which case all data from that cluster is already available for cross-cluster transfers from the beginning, eliminating the need to wait for intra-cluster \texttt{AllGather}.
Figure~\ref{fig:design-pipeline} compares the pipelined and sequential execution of \texttt{AllGahter} in this case.
Thus, \sysname adopts a pipelined execution workflow to overlap independent cross-cluster data transfers and intra-cluster collective steps.

\subsection{Optimality and Tradeoff Analysis}

\head{Optimal cross-cluster data transfer}
Note that the heterogeneous \texttt{SendRecv} makes step \textbf{b)} synchronous across clusters, with the bottleneck determined by the minimum total bandwidth among vendor device groups.
To mitigate this, we optimize both the data transfer volume and bandwidth utilization.
For each collective operation, we estimate the minimum cross-cluster data volume, which is constant for most operations, except for \texttt{Scatter} and \texttt{Gather} for non-root clusters, which scale with cluster size.
Our hierarchical algorithm (Algorithm~\ref{algo:c2c-breakdown}) ensures this volume is minimized to reduce cross-cluster P2P overhead.
Since total bandwidth increases with cluster size, we further balance bandwidth by computing each vendor group’s capacity during initialization and dividing larger groups into subgroups with roughly equal total bandwidth.

\head{System Overhead}
One concern for \sysname's system overhead is the CPU control path, but we consider it acceptable, as CPUs are often underutilized during model training and the control logic is lightweight and implemented in a separate proxy thread.
The vendor-agnostic reduction design sometimes requires additional memory to reduce different data versions, but scratch buffers are inevitable for combining collectives such as \texttt{ReduceScatter}, and \sysname does not require larger scratch buffers than existing approaches.

%% file: fig_design-gdr.tex
\begin{figure}[tbp]
    \centering
    \includegraphics[width=\columnwidth]{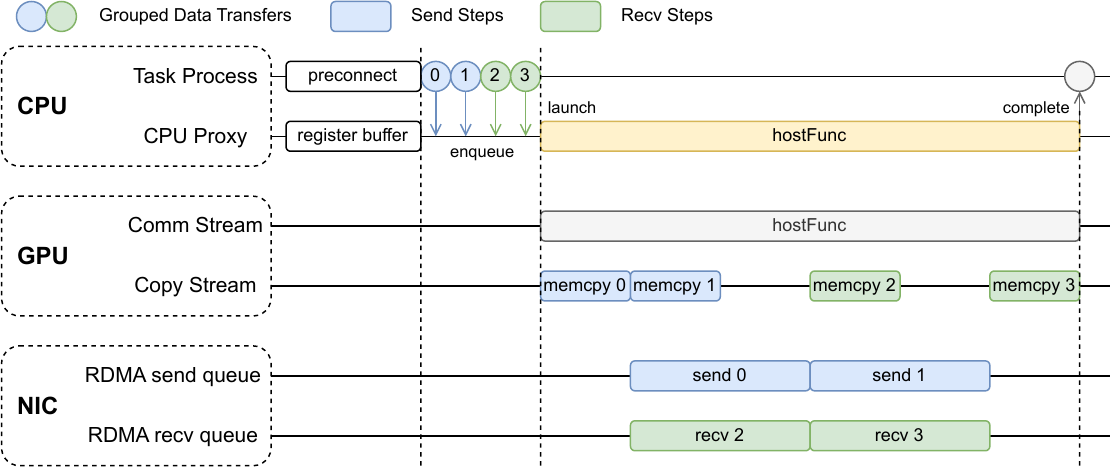}
    \caption{\small P2P Transport for Device Data.}
    \label{fig:design-gdr}
\end{figure}

%% file: fig_design-topo-abst.tex
\begin{figure}[!tbp]
    \centering
    \includegraphics[width=\columnwidth]{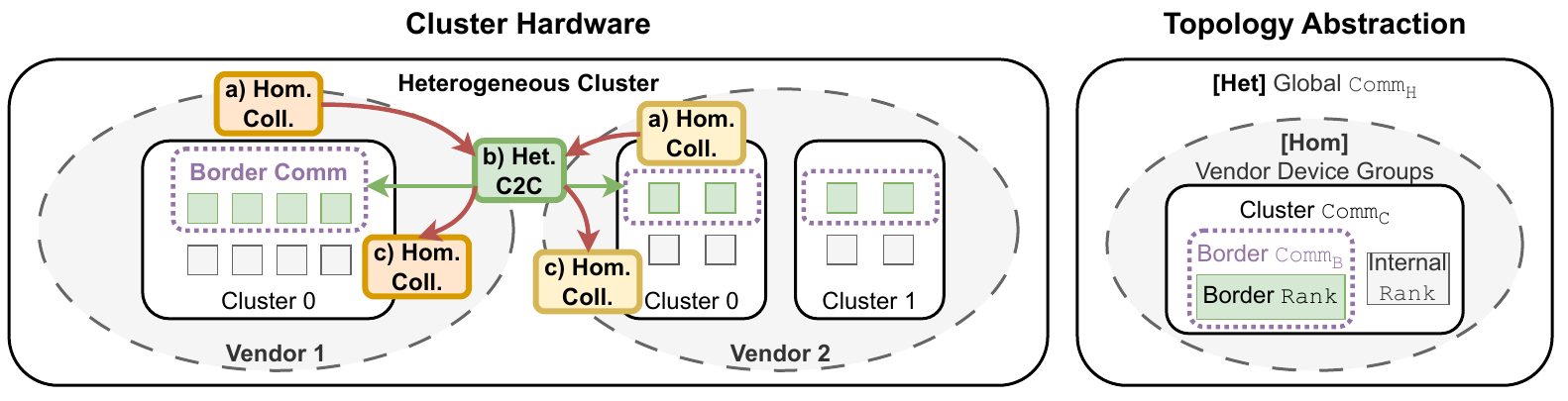}
    \caption{\small Heterogeneous Cluster Topology Abstraction}
    \label{fig:design-topo-abst}
\end{figure}

%% file: tab_design-c2c-primitives.tex
\begin{table}[!btp]
\centering
\resizebox{.9\columnwidth}{!}{\begin{tabular}{
>{\raggedright\arraybackslash}m{1.8cm}|
>{\raggedright\arraybackslash}m{0.8cm}|
>{\raggedright\arraybackslash}m{4cm}|
>{\raggedright\arraybackslash}m{3.5cm}
} \hline
    \textbf{Primitive} & \textbf{Type} & \textbf{Communicator} & \textbf{Custom Params} \\ \hline
    \hline
    \textit{homColl} & HOM & within a $Comm_C$ or $Comm_B$ & send/recv buffer offsets, data length \\ \hline
    \textit{c2cCpy} & HET & $Comm_B$ to $Comm_B$ & send/recv buffer offsets, \#border\_ranks \\ \hline
    \textit{c2cRed} & HET & $Comm_B$ to $Comm_B$ & available bounce buffers, \#border\_ranks \\ \hline
\end{tabular}}
\caption{\small Heterogeneous collective algorithm primitives.}
\label{tab:c2c-primitives}
\end{table}

%% file: fig_design-c2c-algo-prim.tex
\begin{figure*}[!t]
    \centering
    \begin{minipage}[t]{.48\textwidth}
    \includegraphics[width=\columnwidth]{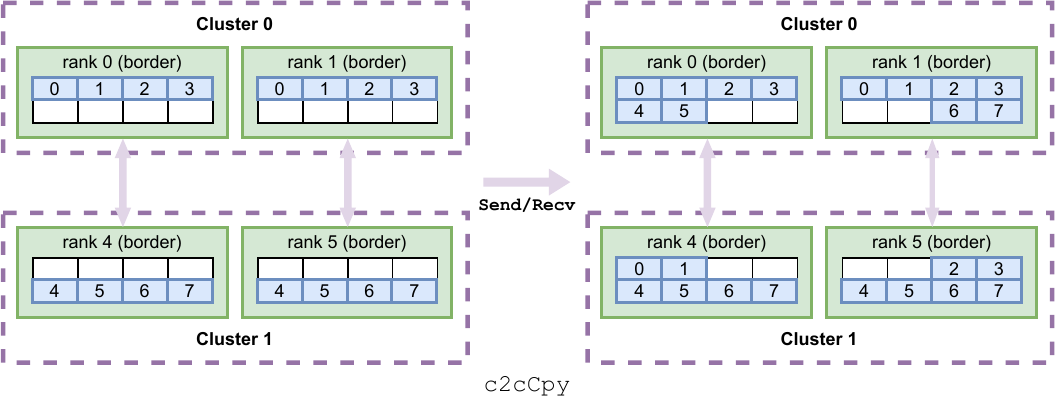}
    \caption{\small Inter-cluster data transfer primitive: \texttt{c2cCpy}}
    \label{fig:design-c2c-cpy}
    \end{minipage}
    \hfill
    \begin{minipage}[t]{.48\textwidth}
    \includegraphics[width=\columnwidth]{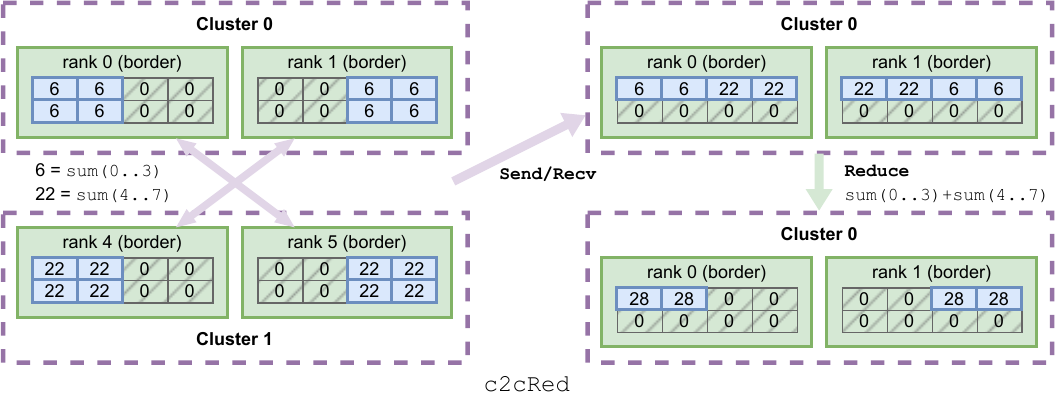}
    \caption{\small Vendor-agnostic reduction primitive: \texttt{c2cRed}}
    \label{fig:design-c2c-red}
    \end{minipage}
\end{figure*}

%% file: algo.tex
\begin{algorithm}[!thb]
  \footnotesize
  \DontPrintSemicolon
  \SetKwInOut{Input}{Params}
  \SetKwInOut{Output}{Output}
  \SetKwData{commh}{$Comm_H$}
  \SetKwData{commc}{$Comm_C$}
  \SetKwData{commb}{$Comm_B$}
  \SetKwData{datai}{data[$i$]}
  \SetKwData{nextcls}{next\_cls}
  \SetKwData{prevcls}{prev\_cls}
  \SetKwData{peercls}{peer\_cls}
  \SetKwData{coll}{$coll$}
  \SetKwData{rank}{$rank$}
  \SetKwIF{If}{Elif}{Else}{if}{then}{else if}{else}{}
  \SetKwFor{For}{for}{do}{}
  \SetKwFor{Foreach}{for each}{do}{}
  
  \SetKwFunction{homcoll}{homColl}
  \SetKwFunction{ctoccpy}{c2cCpy}
  \SetKwFunction{ctocred}{c2cRed}
  \SetKwFunction{gethomcomm}{get\_hom\_comm}

  \SetKwProg{Fn}{Func}{:}{\KwRet}

  \Input{$coll$: collective operation type,\\\commh: global heterogeneous communicator,\\\commc: homogeneous cluster communicator,\\\rank: local rank in the communicator,\\\datai: the $i$th data chunk.}

  \Fn{C2C\_Collective\_Breakdown}{
    \commc $\gets$ \commh.\gethomcomm{rank}\;
    
    \For{$i \gets 1$ \KwTo $n\_start\_loops$}{
      \homcoll{\coll.startColl, \commc, \datai}\;
    }
    
    \For{$i \gets 1$ \KwTo $n\_c2c\_loops$}{
      \If{\coll is non-combining}{
        \ctoccpy{\commc, \commc.\peercls}\;
      }
      
      \Else {
        \ctocred{\commc, \commc.\peercls}\;
      }
    }
    
    \For{$i \gets 1$ \KwTo $n\_end\_loops$}{
      \homcoll{\coll.endColl, \commc, \datai}\;
    }
  }

\caption{\small C2C Collective Breakdown\label{algo:c2c-breakdown}}
\end{algorithm}

%% file: fig_design-pipeline.tex
\begin{figure}[!t]
    \centering
    \includegraphics[width=\columnwidth]{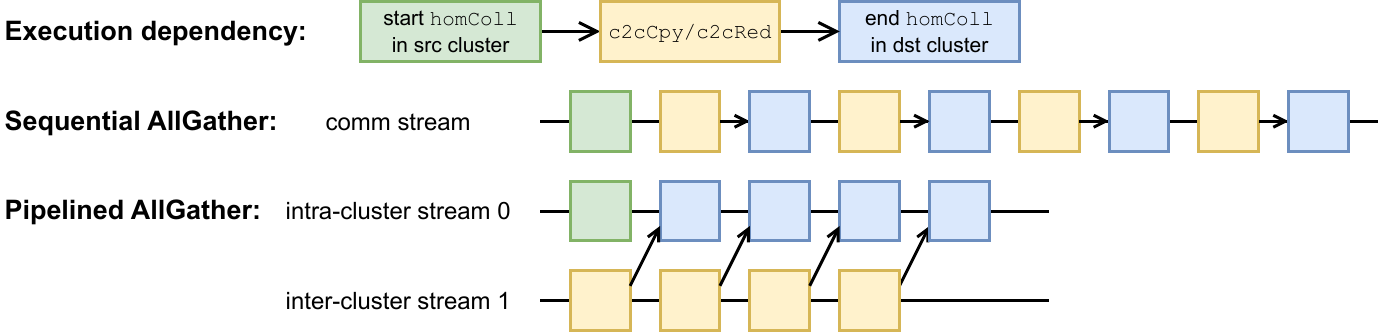}
    \caption{\small Pipelined Collective Algorithm Execution}
    \label{fig:design-pipeline}
\end{figure}

%% file: 5-impl.tex
\section{Implementation\label{sec:impl}}

We implement \sysname with $>30$k LOC in C and Python, including the heterogeneous device-buffer RDMA mechanism, hierarchical topology representation, buffer management for implementing efficient heterogeneous collective communications, several lightweight wrappers of vendor-specific libraries, and PyTorch backend plugin for seamless integration with downstream tasks.

\input{tab_impl-rdma-primitives}

\head{Device APIs}
Using the on-device data path for heterogeneous RDMA requires specific device operations from all vendor hardware.
Table~\ref{tab:rdma-primitives} lists the necessary vendor functions and their definitions, summarized into four types of device APIs: ib\_interface, dev\_mem\_opr, stream\_opr, and host\_func.
The functionality of these APIs is generally supported by all hardware vendors, so we automatically detect the local vendor and redirect these function calls accordingly.
The ib\_interface and device\_mem\_opr ensure hardware access and interaction with device memory and RNICs, and stream\_opr and host\_func provide device-host coordination, including synchronization and device state-checking.

\head{RDMA Transport}
At the bootstrap phase (using a TCP bootstrap network), each rank uses these primitives to get access to available RNICs and register RDMA resources for each heterogeneous connection.
We statically allocate a fixed-size RDMA buffer pool per connection (64M by default, divided into 4M size chunks), and use the host-side asynchronous device-to-device (d2d) memory copy API for moving device memory to RDMA send buffers (or receive buffers to device memory), pipelining d2d memory copying and RDMA transfer for large messages.

\head{Vendor CCL light-weight wrapper}
\sysname integrates each supported vendor's specialized CCL with a lightweight wrapper, and inherits the respective homogeneous communicator (constructed within the heterogeneous communicator for each homogeneous cluster) such that within a homogeneous cluster, \sysname can utilize vendor-optimized collective operation to implement the desired semantics for \sync primitives with better efficiency, without requiring hardware knowledge that may not be publicly available, such as the intra-node interconnect information.

\head{System Integration}\label{subsec:impl-architecture}
PyTorch supports integrating various communication infrastructures with its custom backend feature.
We integrate \sysname's collective communication interfaces into the PyTorch backend plugin feature, such that downstream applications and tasks are unaware of the underlying communication implementation.
Specifically, we extend the \texttt{Work} and \texttt{Backend} classes with a set of APIs as custom C++ extensions~\cite{cpp-ext,third-party-backend}.
With the \sysname PyTorch backend, heterogeneous LLM training can utilize \sysname without changing any application code.

%% file: tab_impl-rdma-primitives.tex
\begin{table}[!b]
\centering
\resizebox{\columnwidth}{!}{\begin{tabular}{>{\raggedright\arraybackslash}m{3.6cm}|>{\raggedright\arraybackslash}m{2cm}|>{\raggedright\arraybackslash}m{5cm}} \hline
    \textbf{Function} & \textbf{Type} & \textbf{Description} \\
    \hline \hline
    \texttt{ibv\_get\_device\_list} & ib\_interface & Get the list of available IB devices on the system \\ \hline
    \texttt{ibv\_open\_device} & ib\_interface & Open IB device, returning a context for future interaction \\ \hline
    \texttt{ibv\_alloc\_pd} & ib\_interface & Create Protection Domain (PD). \\ \hline
    \texttt{ibv\_reg\_mr} & ib\_interface & Register Memory Region (MR). \\ \hline
    \texttt{ibv\_create\_cq} & ib\_interface & Create Completion Queue (CQ). \\ \hline
    \texttt{ibv\_create\_qp} & ib\_interface & Create Queue Pair (QP). \\ \hline
    \texttt{ibv\_modify\_qp} & ib\_interface & Modify QP state. \\ \hline
    \texttt{ibv\_post\_send} \texttt{ibv\_post\_recv} & ib\_interface & Post Work Requests (WRs) to the send/recv queue of a QP. \\ \hline
    \texttt{ibv\_poll\_cq} & ib\_interface & Poll CQ for completed WRs \\ \hline
    \hline
    \texttt{devMalloc/Free} & dev\_mem\_opr & Alloc/free device memory. \\ \hline
    \texttt{d2d/d2h/h2dMemcpy} & dev\_mem\_opr & Device-to-device/device-to-host/host-to-device memory copy. \\ \hline
    \hline
    \texttt{streamCreate/Destroy} & stream\_opr & Create/destroy a stream \\ \hline
    \texttt{streamSync} & stream\_opr & Synchronize streams \\ \hline
    \texttt{streamQuery} & stream\_opr & Query stream state \\ \hline
    \hline
    \texttt{launchHostFunc} & host\_func & Launch a host function on a device stream for maintaining dependency \\ \hline
\end{tabular}}
\caption{\small Device APIs}
\label{tab:rdma-primitives}
\end{table}

%% file: 6-eval.tex
\input{fig_eval-native-wrapper}

\section{Evaluation\label{sec:eval}}

We evaluate \sysname from the following perspectives:
\begin{packedenumerate}
    \item The efficiency of the heterogeneous device-buffer RDMA transport (\sref{subsec:eval-RDMA}). \sysname achieves $> 6 \times$ bandwidth of Gloo in heterogeneous P2P SendRecv operations.
    \item The performance of heterogeneous collective communications (\sref{subsec:eval-coll}). \sysname achieves $> 97\%$ the performance of homogeneous \texttt{AllGather} and $> 70\%$ the performance of homogeneous \texttt{AllReduce} for heterogeneous collective communications.
    \item The efficiency of C2C heterogeneous collective algorithm (Algorithm~\ref{algo:c2c-breakdown}) design (\sref{subsec:eval-micro}). Compared with NCCL, the 2-cluster C2C algorithm achieves $97.4\%$ AllGather bandwidth in a 4-node, 32-GPU cluster. 
    \item The end-to-end performance gain of \sysname in heterogeneous training and serving (\sref{subsec:eval-e2e}). \sysname accelerates the per-step-time for Llama3-3B and Llama3-8B training by $9.1\%$ and $16.9\%$ (\sref{subsec:eval-train}), and reduces TTFT of Qwen-7B serving by $>65\%$ (\sref{subsec:eval-serve}).
\end{packedenumerate}

\head{Testbed Settings}
The evaluation of \sysname involves 4 out of the 8 supported hardware vendors, including a major hardware vendor, NVIDIA A800 (NV in the figures), and three other minor vendors' accelerator cards, which we anonymously refer to as \iluvatar (V1), \cambricon (V2) and \metax (V3), as listed in Table~\ref{tab:eval-testbed}.
The evaluated hardware architectures include both GPGPUs (General-Purpose computing on Graphics Processing Units) and ASIC (Application-Specific Integrated Circuits).
\input{tab_eval-testbed}

\head{CCL Baselines}
We compare \sysname with vendor-provided native CCLs (NCCL, \accl, \bccl, and \cccl) and a host-forwarding approach implemented with Gloo.
Specifically, the CPU bounce buffers in Gloo are transferred via RDMA transport in our baseline.

\head{Vendor CCL Wrapper Efficiency}
Figure~\ref{fig:eval-native-wrapper} compares the algorithm bandwidth of vendor CCLs and the \sysname wrapper with NCCL-style perf-test of SendRecv, AllGather, and AllReduce operations.
The left and right sub-figures represent NCCL and \accl, and the solid and dashed lines represent the wrapper and the native CCLs.
\sysname achieves $98\%\sim99\%$ bandwidth of the native CCLs for arbitrary message sizes, which validates that \sysname wrapper overhead is negligible.
For the rest of the evaluation, we only show the performance of \sysname wrappers as the vendor CCL performance baselines.

\subsection{Benchmark Evaluation\label{subsec:eval-benchmark}}

\subsubsection{Heterogeneous P2P Performance\label{subsec:eval-RDMA}}

Figure~\ref{fig:eval-p2p} shows the device-buffer RDMA transport performance of \sysname.
We compare the SendRecv bandwidth of \sysname with vendor CCLs (for homogeneous peers) and with Gloo (for CPU-forwarding between heterogeneous peers, NVIDIA and \metax).
Figure~\ref{fig:eval-p2p} shows that \sysname achieves $> 6 \times$ higher bandwidth than the existing heterogeneous communication framework Gloo.

We use linear regression to synthesize the latency and bandwidth according to the $\alpha$-$\beta$ cost model\footnote{The $R^2$ (R-squared) value indicates precisions higher than 0.999.} for P2P transport.
We label the synthesized latency and bandwidth on the horizontal dotted lines in Figure~\ref{fig:eval-p2p}.
Under asymmetric hardware bandwidth settings, \sysname achieves $13 \sim 45\%$ higher bandwidth than the lower homogeneous P2P bandwidth between the two hardware vendors.
For symmetric hardware bandwidth settings (using \sysname's device-buffer RDMA transport between two NVIDIA GPUs), \sysname achieves $10.4 \%$ higher bandwidth than NCCL SendRecv operation.
The synthesized latency cost of \sysname is $1.2 \sim 2.4 \times$ vendor-specialized libraries ($0.05 \sim 0.18ms$ VS. $0.10 \sim 0.40ms$), but still significantly lower than Gloo ($1.73ms$).

\input{fig_eval-benchmark}

\input{fig_eval-e2e}

\subsubsection{Collective Communication Benchmarks\label{subsec:eval-coll}}

Figure~\ref{fig:eval-ag} and Figure~\ref{fig:eval-ar} show the collective communication performance of \sysname for \texttt{AllGather} and \texttt{AllReduce}, which are the most commonly used non-combining and combining collectives in LLM training tasks.
Different heterogeneous hardware combinations using \sysname (NVIDIA A800 with hardware from each of the three minor vendors, plus a combination of two minor vendors, \cambricon and \metax) are compared with vendor CCLs in their respective homogeneous environments, all using a 2-node setup.
The heterogeneous \texttt{AllGather} in \sysname achieves $85.7 \sim 97.8\%$ of the bandwidth of the slower homogeneous \texttt{AllGather} implementation from the two vendors, as the performance of the slower hardware inevitably becomes the bottleneck.
For \texttt{AllReduce}, as we trade more intra-cluster communication for compatibility (\sref{subsubsec:design-2-c2c-prim}), the bandwidth can reach up to $70.8\%$ of that of the slower vendor implementation.

\subsubsection{Algorithm Design Microbenchmarks\label{subsec:eval-micro}}

To assess the \sysname collective algorithm design regardless of hardware capability discrepancies, we evaluate the C2C algorithm against NCCL native implementation using 4 NVIDIA server nodes, each containing 8 A800 GPUs.
In Figure~\ref{fig:eval-c2c}, the 2+2 C2C lines denote the performance of running Algorithm~\ref{algo:c2c-breakdown} on a hierarchical topology of two 2-node clusters, while the 4-node native lines denote the performance of native NCCL in the same cluster.
\sysname achieves $97.4\%$ and $59.1\%$ bandwidth of NCCL for \texttt{AllGather} and \texttt{AllReduce} operations, respectively.

For the last benchmark, we test the compatibility and scalability of \sysname to multi-NIC hardware.
Figure~\ref{fig:eval-multi-nic} shows the \texttt{AllGather} and \texttt{AllReduce} performance of \sysname in our 4-node NVIDIA A800 environment, each node equipped with 8 GPUs per node and using 1/2/4/8 RNICs per node.
The collective bandwidth grows proportionally to the number of NICs in use, validating the compatibility and high hardware utilization of \sysname.

\subsection{End-to-end Evaluation\label{subsec:eval-e2e}}

\subsubsection{LLM Training Performance\label{subsec:eval-train}}

\head{Communication Speedup}
First, we evaluate the end-to-end training speedup of \sysname compared to the host-forwarding approach.
Detailed setups are listed in Appendix~\ref{append:e2e-setup} Table~\ref{tab:eval-e2e-setup-perf}.
Given hardware differences, we adopt asymmetric parallel strategies~\cite{park2020hetpipe} to match computational capabilities, and reduce 1 PP layer at the start and end nodes for the extra embedding and loss computation.
This is the best heterogeneous training strategy we know of for this testbed.
\sysname's communication speedup should be effective regardless of parallel strategies, and optimizing heterogeneous parallel training strategies is beyond the scope of this research.
Figure~\ref{fig:eval-e2e-perf} shows that \sysname can improve Llama3-3B and Llama3-8B training throughput by 9.98\% and 20.38\%. 
The performance gain in end-to-end training evaluation is less significant than the benchmark because computation dominates the training time in smaller models and clusters.
As communication becomes more dominant in larger models and clusters, we expect \sysname to yield greater speedup as the training scale grows.

\head{Hardware Scalability}
Figure~\ref{fig:eval-e2e-scale} demonstrates the training performance comparison of \sysname and native vendor CCLs in another testbed consisting of NVIDIA and \metax hardware.
\sysname effectively utilizes heterogeneous computational resources, which were previously unavailable due to the lack of a heterogeneous collective communication library.
We train the Llama3-8B model using various parallel strategies in 5 different setups: two homogeneous (two NVIDIA A800 servers, two HW-3 servers) and three heterogeneous (1 A800 + 1 HW-1, 2 A800 + 2 HW-1, 4 A800 + 4 HW-1).
Despite hardware heterogeneity, the per-step time using two heterogeneous servers increases by only $7.6\%$ compared to using two homogeneous NVIDIA servers.
The training throughput improves by $56\%$ and $64\%$ when using a 4-node heterogeneous cluster compared to 2-node homogeneous NVIDIA or \metax clusters, respectively.
The 8-node heterogeneous cluster additionally improves training throughput by up to $51\%$ compared to the 4-node setup.

\subsubsection{LLM Serving Performance\label{subsec:eval-serve}}

Lastly, we evaluate the LLM serving performance of \sysname, demonstrating another promising use case of mixed-vendor cluster deployment.
We adopt the recent prefill-decode disaggregation approach~\cite{zhong2024distserve} to accommodate the different capabilities of heterogeneous hardware, as the prefill phase is more computation-intensive and the decode phase is more memory-intensive.
Detailed evaluation setup is as listed in Appendix~\ref{append:serve-setup} Table~\ref{tab:eval-serving-setup}.
We compare the mean, median, and 99-percentile (99P) of time-to-first-token (TTFT) (Figure~\ref{fig:eval-e2e-serve}) and end-to-end output and total token throughput of 100 requests (Figure~\ref{fig:eval-e2e-tp}) of the Qwen2-7B model serving, where the data transfer across prefill and decode phases is carried out by NCCL (for the NVIDIA homogeneous setup), host-forwarding (for both setups), or \sysname (treating both setups as heterogeneous).
\sysname reduces $65\%$ TTFT and improves $19\%$ output token throughput compared with the host-forwarding data transfer approach.
Compared with NCCL, using \sysname on two NVIDIA A800 GPUs yields slightly longer TTFT but outperforms NCCL in output and total token throughput.
In addition, in the heterogeneous hardware settings, compared with using NCCL for two NVIDIA A800 GPUs, using a combination of NVIDIA and \metax hardware induces less than $30\%$ longer TTFT and decreases throughput by $33\%$, which is bottlenecked by the hardware processing capability rather than communication.
This indicates that coordinating heterogeneous serving strategies with \sysname communication hides a large portion of hardware differences across vendors, leading to more efficient hardware utilization.

%% file: fig_eval-native-wrapper.tex
\begin{figure*}
\centering
\begin{minipage}[t]{.66\textwidth}
    \resizebox{\columnwidth}{.3\columnwidth}{
    \includegraphics[width=\columnwidth]{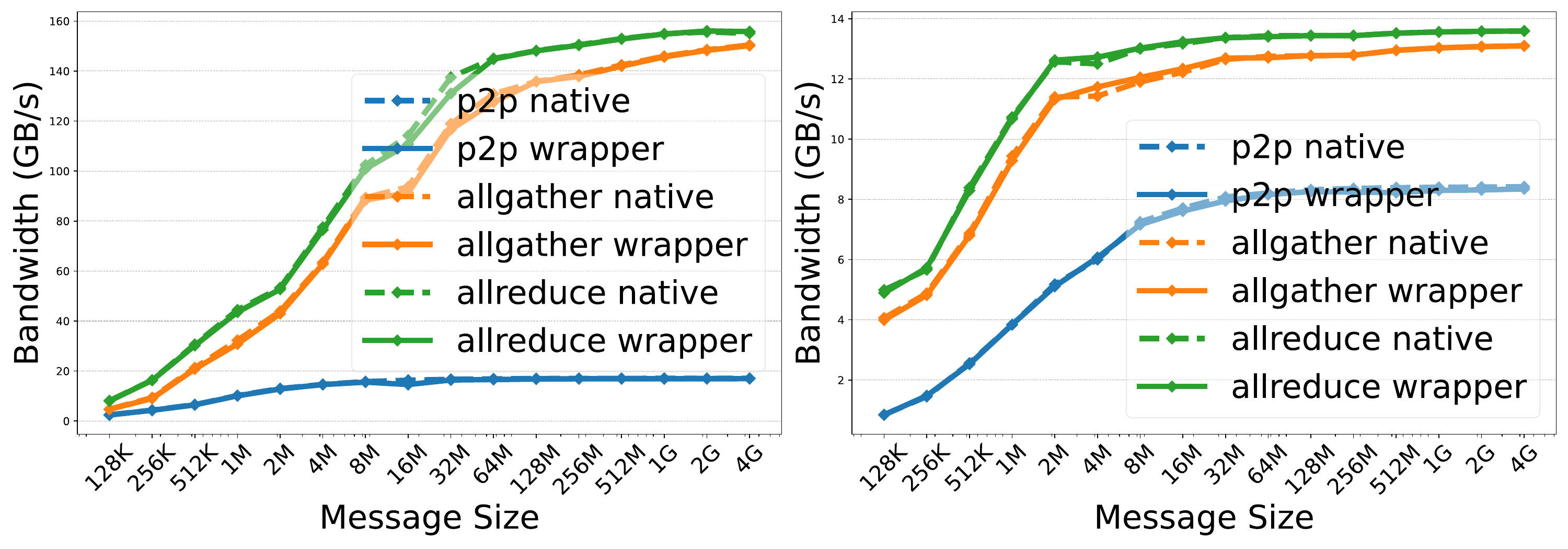}
    }
    \caption{\small \sysname vendor CCL wrapper performance}
    \label{fig:eval-native-wrapper}
\end{minipage}
    \hfill
\begin{minipage}[t]{.33\textwidth}
    \resizebox{\columnwidth}{.6\columnwidth}{
    \includegraphics[width=\columnwidth]{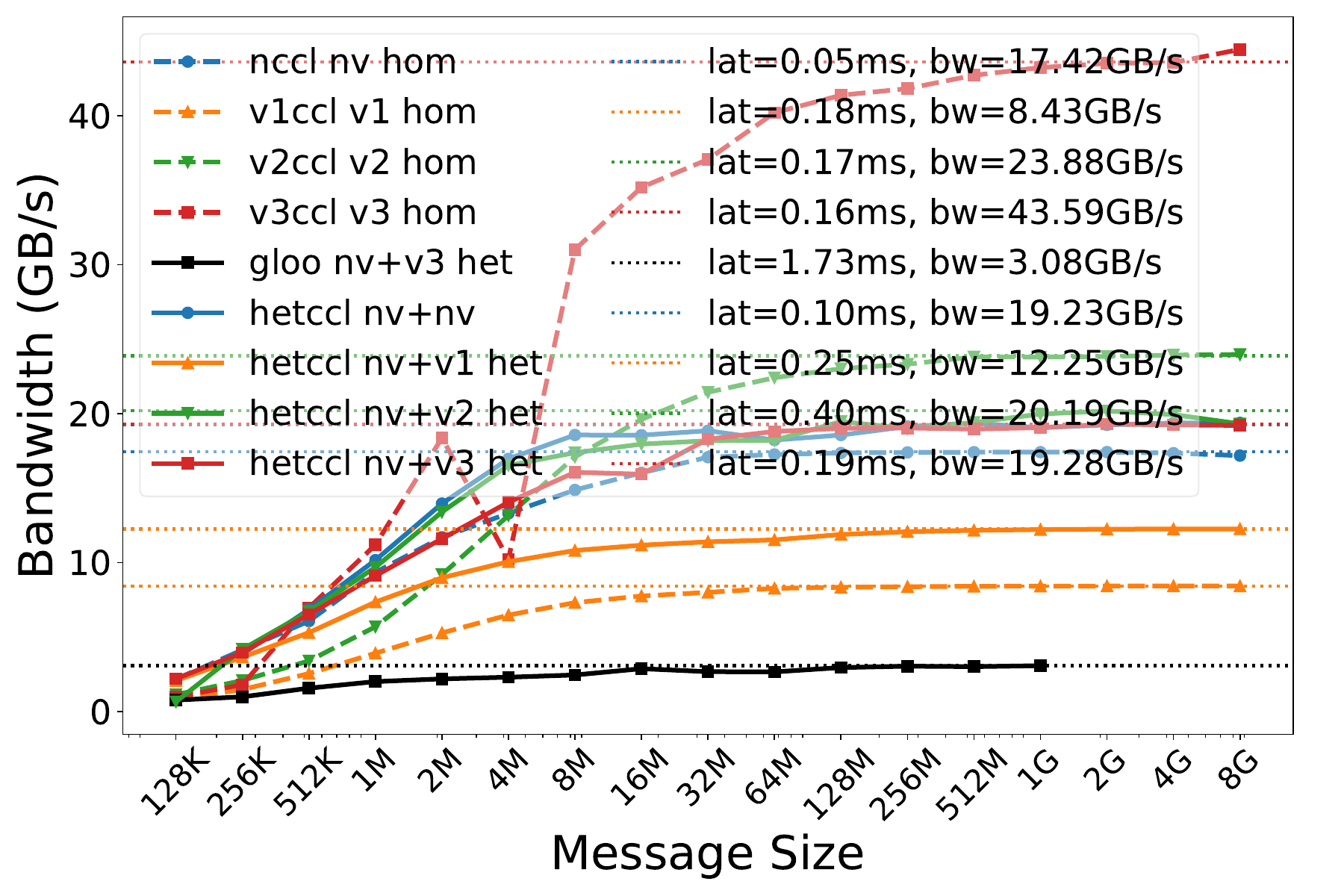}
    }
    \caption{\small P2P Data Transfer Performance}
    \label{fig:eval-p2p}
\end{minipage}
\end{figure*}

%% file: tab_eval-testbed.tex
\begin{table}[!b]
    \centering
    \resizebox{\columnwidth}{!}{
    \begin{tabular}{c|c|c|c|c|c|c} \hline
        \textbf{Vendor} & \textbf{Hardware} & \textbf{\#Node} & \textbf{\#dev/node} & \textbf{RNIC BW} & \textbf{Code} & \textbf{Topo}  \\
        \hline\hline
        NVIDIA     & A800 (GPGPU) & 4      & 8          & 8x200Gbps & Y & Y           \\
        \hline
        \iluvatar  & HW-1 (GPGPU) & 2      & 16         & 1x100Gbps & Y & N           \\
        \hline
        \cambricon & HW-2 (ASIC)  & 2      & 8          & 8x400Gbps & N & N           \\
        \hline
        \metax     & HW-3 (GPGPU) & 4      & 8          & 8x400Gbps & N & N           \\
        \hline
    \end{tabular}
    }
    \caption{\small \sysname evaluation testbed settings}
    \label{tab:eval-testbed}
\end{table}

%% file: fig_eval-benchmark.tex
\begin{figure*}[!t]
\centering
    \begin{minipage}[t]{.24\textwidth}
        \resizebox{\columnwidth}{.6\columnwidth}{
        \includegraphics[width=\columnwidth]{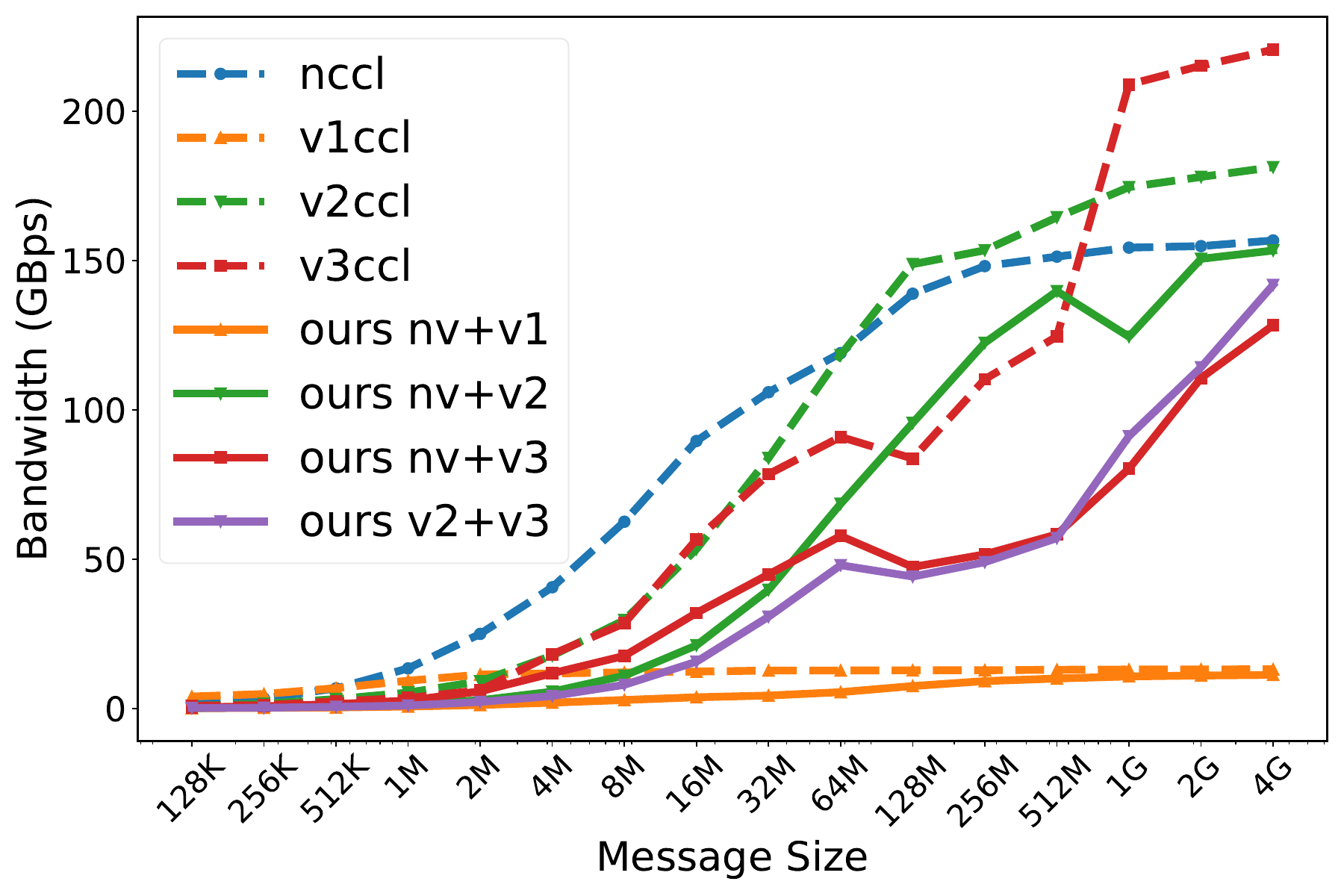}
        }
        \caption{
        \small Heterogeneous \texttt{AllGather} Performance
        }
        \label{fig:eval-ag}
    \end{minipage}
    \hfill
    \begin{minipage}[t]{.24\textwidth}
        \resizebox{\columnwidth}{.6\columnwidth}{
        \includegraphics[width=\columnwidth]{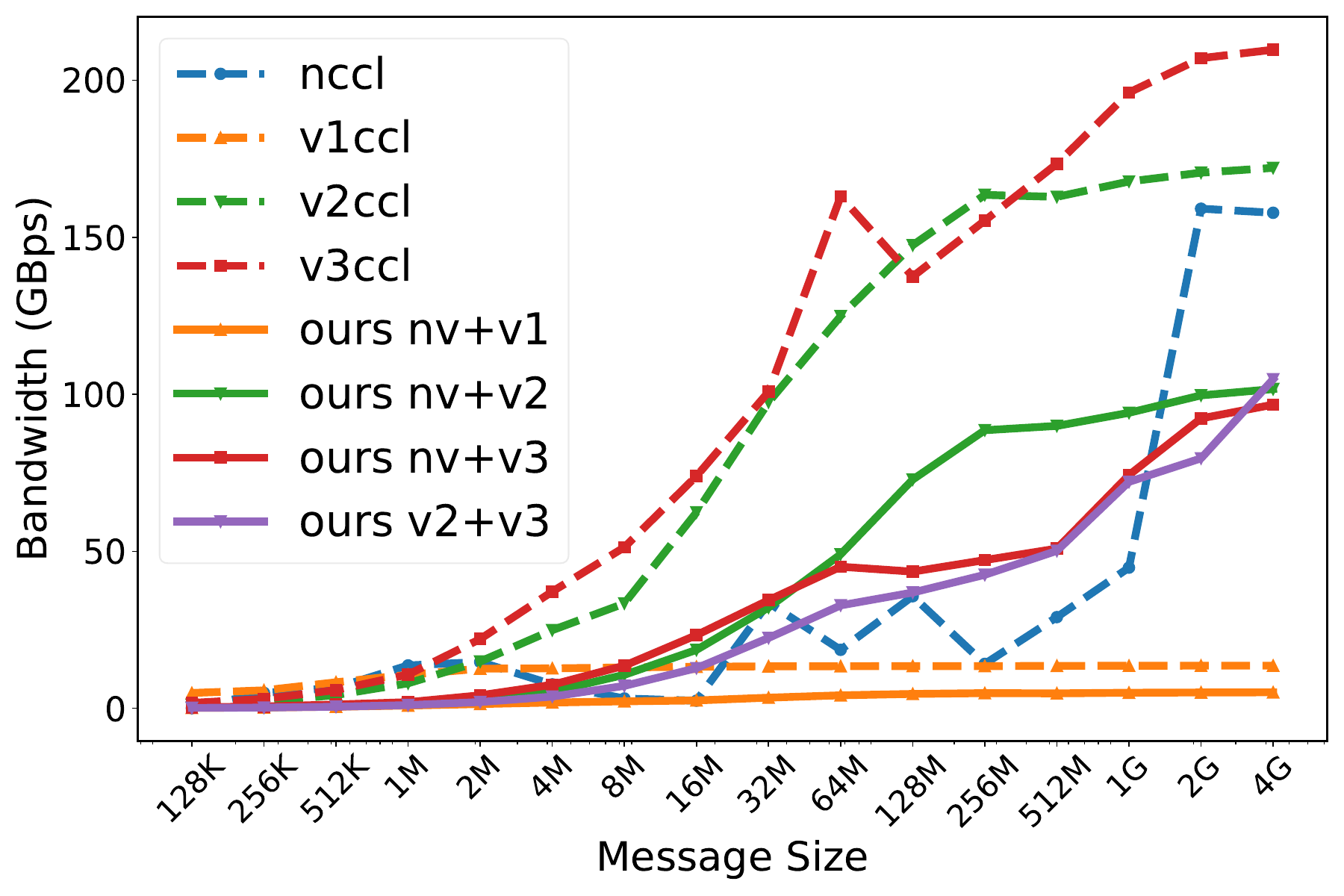}
        }
        \caption{
        \small Heterogeneous \texttt{AllReduce} Performance
        }
        \label{fig:eval-ar}
    \end{minipage}
\hfill
\begin{minipage}[t]{.24\textwidth}
    \centering
    \resizebox{\columnwidth}{.6\columnwidth}{
        \includegraphics[width=\columnwidth]{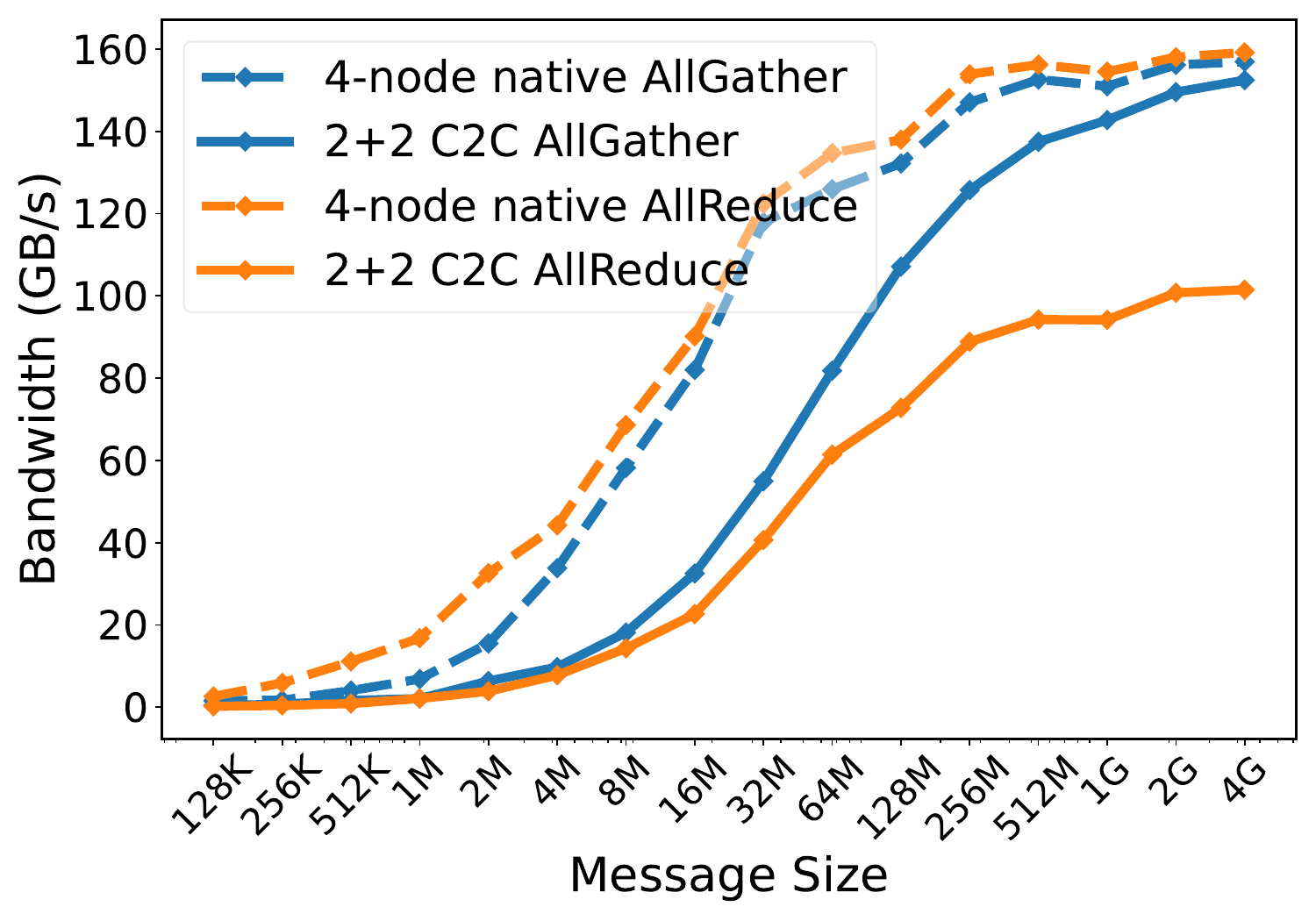}
    }
    \caption{\small Efficiency of C2C collective breakdown}
    \label{fig:eval-c2c}
\end{minipage}
\hfill
\begin{minipage}[t]{.24\textwidth}
    \centering
    \resizebox{\columnwidth}{.6\columnwidth}{
        \includegraphics[width=\columnwidth]{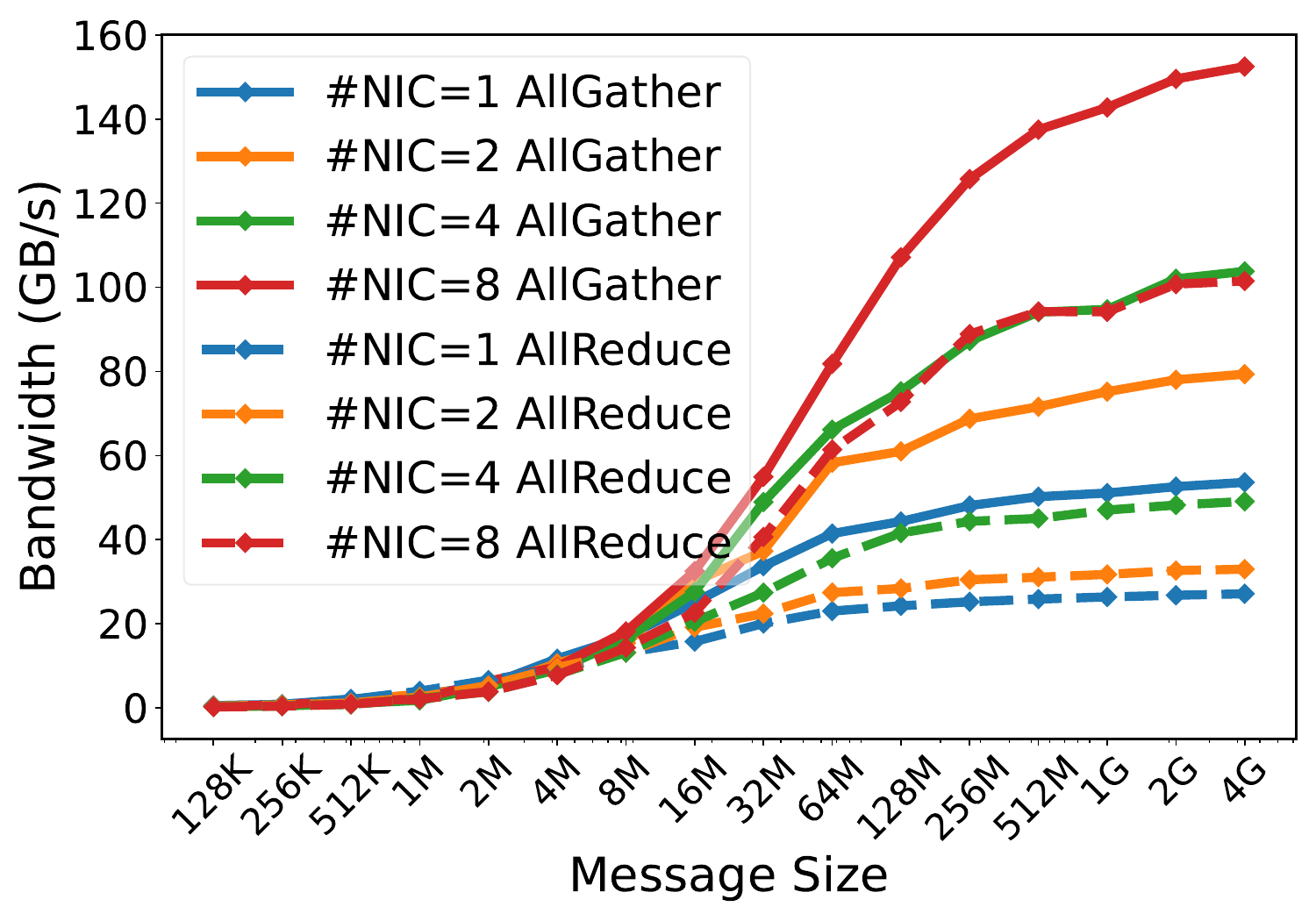}
    }
    \caption{\small Compatibility with multi-NIC hardware}
    \label{fig:eval-multi-nic}
\end{minipage}
\end{figure*}

%% file: fig_eval-e2e.tex
\begin{figure*}[!t]
\centering
\begin{minipage}[t]{.24\textwidth}
    \centering
    \resizebox{\columnwidth}{.6\columnwidth}{
    \includegraphics[width=\columnwidth]{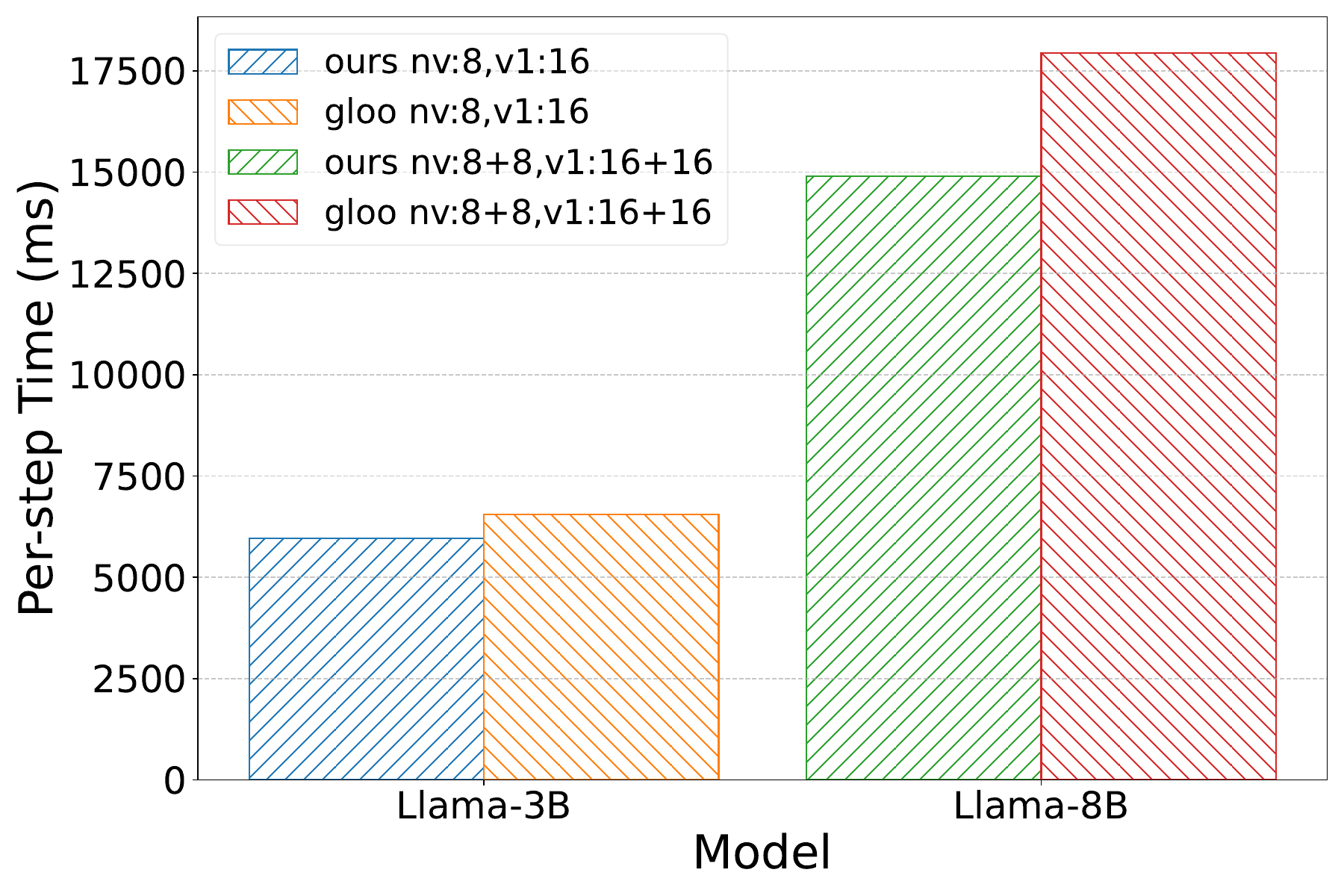}
    }
    \caption{\small End-to-end speed up of \sysname over Gloo}
    \label{fig:eval-e2e-perf}
\end{minipage}
\hfill
\begin{minipage}[t]{.24\textwidth}
    \centering
    \resizebox{\columnwidth}{.6\columnwidth}{
    \includegraphics[width=\columnwidth]{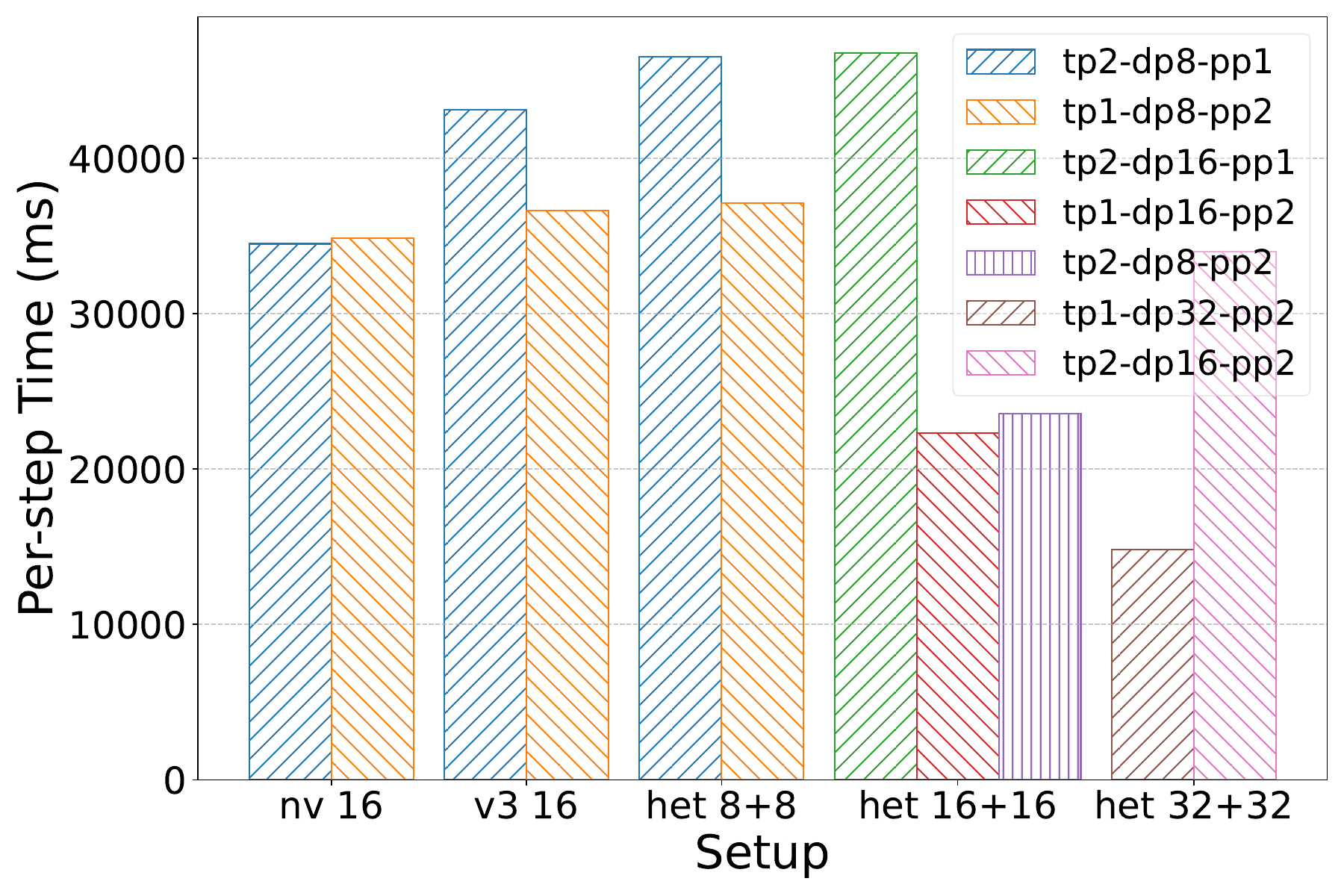}
    }
    \caption{\small End-to-end training scalability of \sysname}
    \label{fig:eval-e2e-scale}
\end{minipage}
\hfill
\begin{minipage}[t]{.24\textwidth}
    \centering
    \resizebox{\columnwidth}{.6\columnwidth}{
    \includegraphics[width=\columnwidth]{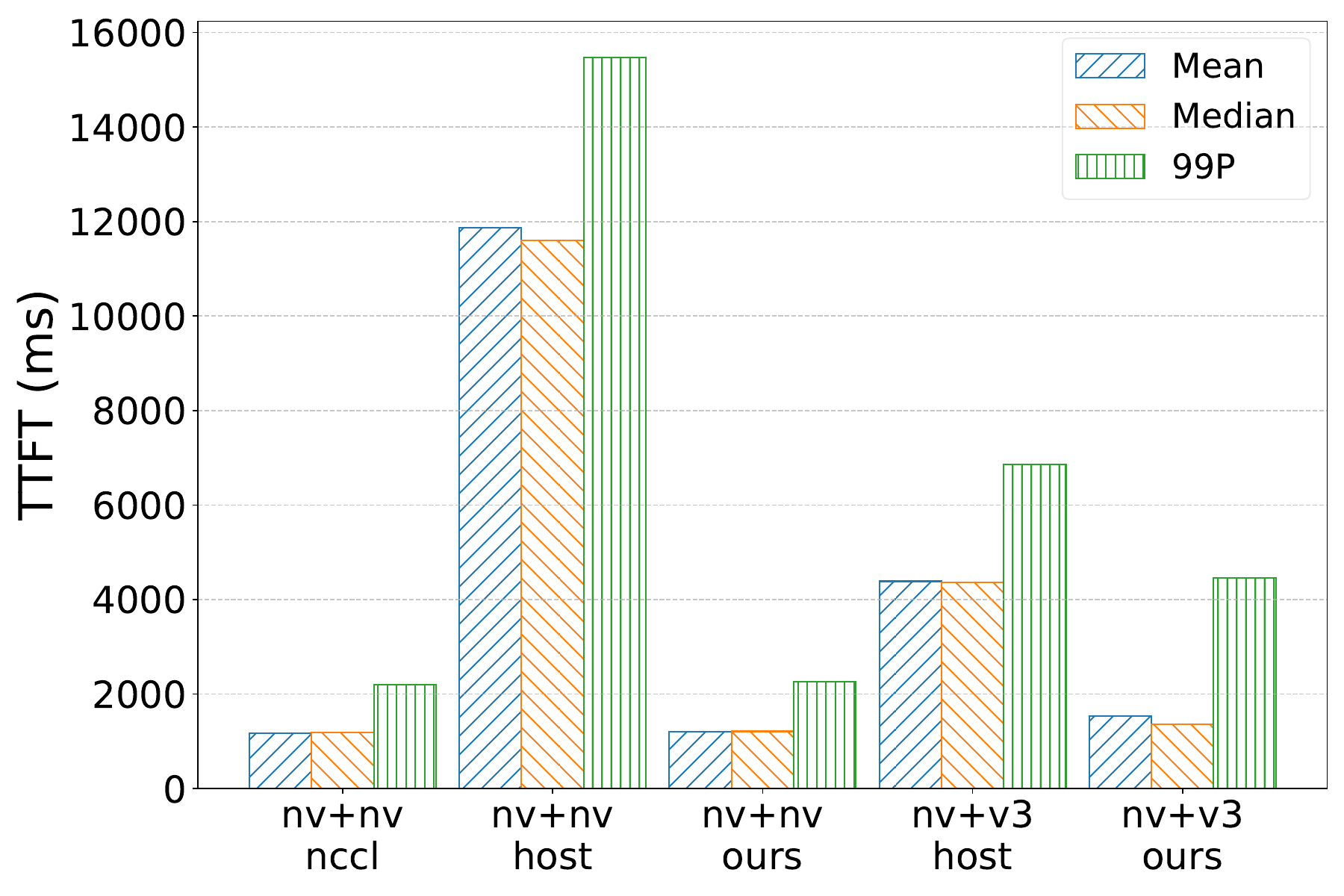}
    }
    \caption{\small End-to-end serving latency performance}
    \label{fig:eval-e2e-serve}
\end{minipage}
\hfill
\begin{minipage}[t]{.24\textwidth}
    \centering
    \resizebox{\columnwidth}{.6\columnwidth}{
    \includegraphics[width=\columnwidth]{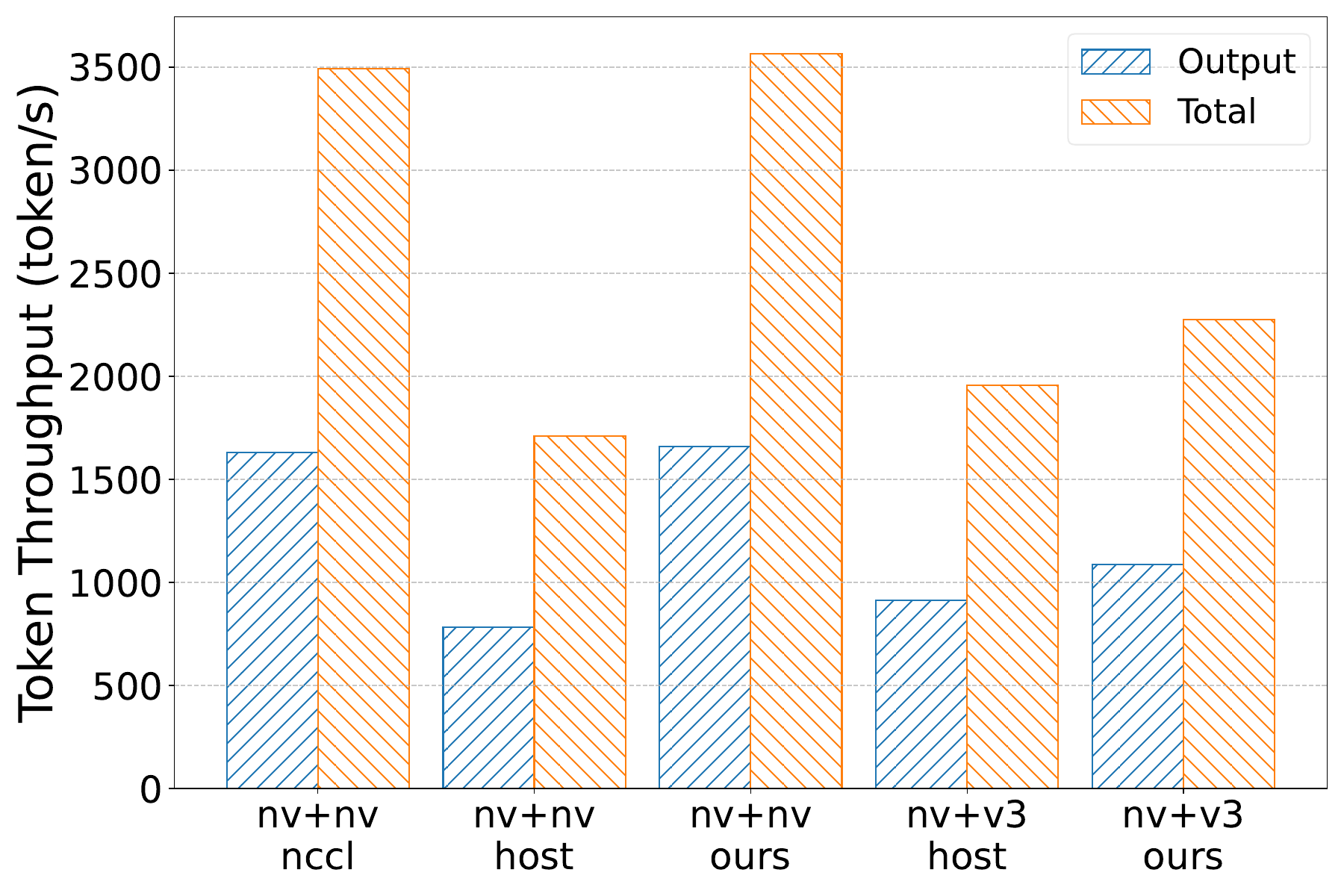}
    }
    \caption{\small End-to-end serving throughput performance}
    \label{fig:eval-e2e-tp}
\end{minipage}
\end{figure*}

%% file: 7-conclusion.tex
\section{Related Works\label{sec:related}}

\head{Heterogeneous Communication Frameworks}
Gloo~\cite{gloo} and OpenMPI~\cite{chen2023mpi} use the CPU-forwarding mechanism (Figure~\ref{fig:motivation-data-path}), incurring high data-path overhead compared to our on-device approach.
The Unified Collective Communication Library (UCC~\cite{DBLP:conf/hoti/VenkataPLBALBDS24}), built on Unified Communication X (UCX~\cite{ucx,openucx-home,openucx-git}), employs a component-based design with team layers (TLs) (e.g., UCX, CUDA, NCCL) for transport abstraction. However, UCC also relies on CPU-forwarding for cross-device communication, facing the same CPU and PCIe bottlenecks as Gloo and OpenMPI.

\head{Collective Algorithm Optimizations}
Several works leverage MSCCL~\cite{cai2021synthesizing} to decompose operations into P2P send/recv primitives.
TACCL~\cite{shah2023taccl} enhances algorithm search with user-input communication sketches, while TCCL~\cite{kim2024tccl} focuses on the path-finding and congestion issues in PCIe GPU clusters.
TE-CCL~\cite{liu2024rethinking} models communication as a multi-commodity flow problem with MILP encoding. 
ForestColl~\cite{zhao2024forestcoll} employs a spanning tree-based approach with polynomial solving time but is limited to tree-based algorithms and optimizes only for bandwidth.
Note that the \textit{heterogeneous} support claimed by their mutual execution backend, MSCCL~\cite{cai2021synthesizing}, refers to supporting deployment of its algorithms on NVIDIA or AMD GPUs, but each deployment itself has to be homogeneous, \ie not applicable to a cluster containing both vendors' hardware, which keeps all these approaches from generalizing to heterogeneous clusters.

\head{Portable Programming Models}
Programming models such as Triton~\cite{tillet2019triton,zheng2025triton}, SYCL~\cite{sycl} and OpenACC~\cite{openacc} are portable across GPUs from multiple vendors.
However, they are designed to write portable kernels for computation tasks such as matrix multiplication or convolution, which lack the necessary interfaces such as transport resource management and IB operations, to implement kernel-based communication (\eg device RDMA without CPU intervention).

\head{Heterogeneous Training}
Existing works such as Whale~\cite{jia2022whale}, Pipette~\cite{pipette}, and Metis~\cite{metis} explore parallel strategies such as tensor parallelism, data parallelism, and pipeline parallelism, relying on collective communication operations like AllReduce, Broadcast, and AllGather to synchronize data across devices. This line of research is orthogonal to our communication optimizations.
Regardless of the parallel strategy, heterogeneous training requires high communication efficiency to prevent bottlenecks. \sysname tackles this challenge, enabling scalable and efficient training across multi-vendor GPU environments.

\section{Discussion\label{subsec:discuss}}

\head{Device-centric RDMA}
\sysname's device RDMA depends on a CPU-centric control logic.
Implementing a completely device-logic collective faces the following problems, which we seek to solve in future works:
1) \textit{Lack in programmability:} Device programming libraries (CUDA, ROCm) lack universal abstractions/interfaces to use kernel threads for RDMA control paths. 
2) \textit{Performance degradation:} Extending existing frameworks with RDMA primitives to support entirely kernel-based heterogeneous send/recv usually comes with performance degradation and limitations in the algorithm search space.

\head{Intra- and inter-cluster coordinated algorithm design}
\sysname currently relies on vendors to provide high performance homogeneous collective communications.
For future improvements, we seek to extend the algorithm encoding and search engine to consider intra-cluster as well as inter-cluster hardware features and topologies to reduce vendor-provided functionality requirements and optimize heterogeneous collective algorithms with a more comprehensive approach.

\head{Failure handling}
\sysname currently does not provide additional failure handling mechanisms.
The vendor-CCL wrapper component in \sysname inherits the native failure-handling capability provided in these third-party libraries, and \sysname further provides NCCL-style logs for the heterogeneous device data P2P transport for failure root cause analysis.

\section{Conclusion}

We propose \sysname, a novel heterogeneous collective communication framework, which implements heterogeneous device-buffer RDMA and utilizes homogeneous high performance collective operations to construct efficient heterogeneous collective algorithms in multi-vendor clusters.
We implement and evaluate \sysname with 4 hardware vendors, which achieves over $97\%$ and $70\%$ of the performance of homogeneous \texttt{AllGather} and \texttt{AllReduce} collectives.
\sysname is integrated into PyTorch using the customized backend feature, which accelerates LLM training jobs by up to $16.9\%$.

%% file: appendix-nsdi.tex
\input{tab_design-collectives-breakdown}

\input{fig_design-c2c-algo-ag}

\input{fig_design-c2c-algo-ar}

\section{Hierarchical Algorithm for Heterogeneous Collectives\label{append:coll-algo}}

In this section, we present the detailed collective breakdown logic of \texttt{AllGather} and \texttt{AllReduce} in Figure~\ref{fig:design-c2c-ag} and \ref{fig:design-c2c-ar}.
We demonstrate the operations in a heterogeneous environment with two types of hardware (divided into Cluster 0 and Cluster 1, respectively), each with 4 devices (4 ranks) and 2 RNICs (2 border ranks).
The precise semantics of \texttt{AllGather} and \texttt{AllReduce} operations are constructed with our cluster-level primitives (\sref{subsubsec:design-2-c2c-prim}).
The \texttt{start}, \texttt{c2c}, and \texttt{end} operations of Algorithm~\ref{algo:c2c-breakdown} for other collective operations are as listed in Table~\ref{tab:c2c-primitives}.

\section{End-to-end Evaluation Setups\label{append:e2e-setup}}

\subsection{Training Evaluation Setups\label{append:train-setup}}

\input{tab_eval-e2e-setup}

\head{Communication Speedup}
Table~\ref{tab:eval-e2e-setup-perf} lists the evaluation setups for communication comparison between \sysname and Gloo.
In Setup 1, we train a Llama3-3B model in a mixed-vendor environment with 1 NVIDIA 8-GPU server and 1 \iluvatar 16-GPU server.
Setup 2 trains a Llama3-8B model in a mixed-vendor environment with 2 NVIDIA 8-GPU servers and 2 \iluvatar 16-GPU servers.
We adopt pipeline parallelism across the two vendor hardware groups, with the pipeline layers divided according to each hardware's computation capability.
Then we adopt data parallelism within the group (DP equals the number of GPUs per PP group).
The same model and parallel strategies are executed with \sysname and Gloo as communication backends, respectively, to evaluate the end-to-end impact of communication efficiency.

\head{Hardware Scalability}
Table~\ref{tab:eval-e2e-setup-scale} lists the setups for hardware scalability evaluation of \sysname.
We train the same Llama3-8B model in both homogeneous and heterogeneous clusters of various scales.
For NVIDIA and \metax homogeneous clusters, we train with their native CCLs, respectively, and for heterogeneous setups, we train with \sysname.
For 2-node homogeneous and heterogeneous setups (Setup 3, 4, and 5), we test two different parallel strategies, namely $[TP,DP,PP]=[2,8,1]$ and $[TP,DP,PP]=[1,8,2]$.
For 4-node and 8-node heterogeneous setups (Setup 6 and 7), we extend the previous parallel strategies with $DP$ groups or $PP$ groups to evaluate \sysname's ability to utilize heterogeneous computation resources.

\subsection{Serving Evaluation Setups\label{append:serve-setup}}

\input{tab_eval-serving-setup}
Table~\ref{tab:eval-serving-setup} lists the end-to-end evaluation setups for LLM serving scenario.
In both setups, we put the prefill phase on NVIDIA hardware and the decode phase on the other hardware.
The Qwen2-7B serving is implemented with \texttt{vllm}, adapted to transfer the KV cache between prefill and decode phase with various communication backends (NCCL, host-forwarding and \sysname).

%% file: tab_design-collectives-breakdown.tex
\begin{table*}[!b]
\centering
\resizebox{\textwidth}{!}{\begin{tabular}{
|>{\raggedright\arraybackslash}m{2.5cm}  
|>{\raggedright\arraybackslash}m{3.5cm}  
|>{\raggedright\arraybackslash}m{4.5cm}  
|>{\raggedright\arraybackslash}m{3cm}  
|>{\raggedright\arraybackslash}m{3cm}  
|>{\raggedright\arraybackslash}m{3.5cm}  
|}
    \hline
    \textbf{Heterogeneous Collective} & \textbf{start $homColl$} & \textbf{C2C primitive} & \textbf{C2C total send volume} & \textbf{C2C total recv volume} & \textbf{end $homColl$} \\
    \hline
    \texttt{AllReduceH} & \texttt{Reduce} $/$ \texttt{ReduceScatter} & b1)~\texttt{c2cRed: ReduceScatter} b2)~\texttt{c2cCpy} & $\approx 2*n*(C-1)/C$ & $\approx 2*n*(C-1)/C$ & \texttt{Bcast} $/$ \texttt{No-op} \\
    \hline
    \texttt{AllGatherH} & \texttt{AllGather} & \texttt{c2cCpy} & $\approx G * n$ & $\approx G * n$ & \texttt{Bcast} \\
    \hline
    \texttt{ReduceScatterH} & \texttt{Reduce} & \texttt{c2cRed: ReduceScatter} & $(G-N)*n$ & $(C-1)*N*n$ & \texttt{ReduceScatter} \\
    \hline
    \texttt{BcastH} & \texttt{Bcast} & \texttt{c2cCpy} & $n$ (root) & $n$ (non-root) & \texttt{Bcast} \\
    \hline
    \texttt{ReduceH} & \texttt{Reduce} & \texttt{c2cRed: ReduceScatter} & $n$ (non-root) & $n$ (root) & \texttt{Reduce} \\
    \hline
    \texttt{GatherH} & \texttt{Gather} & \texttt{c2cCpy} & $N * n$ (non-root) & $(G-N) * n$ (root) & \texttt{Gather} \\
    \hline
    \texttt{ScatterH} & \texttt{Scatter} & \texttt{c2cCpy} & $(G-N)*n$ (root) & $N*n$ (non-root) & \texttt{Scatter} \\
    \hline
    \texttt{AllToAllH} & \texttt{AllToAll} $/$ \texttt{sendrecv} & \texttt{c2cCpy} & $(G-N)*n$ & $(G-N)*n$ & \texttt{SendRecv}/ \texttt{Scatter} \\
    \hline
\end{tabular}}
\small We use $C$ for the number of clusters and $G$ for the total number of ranks in the heterogeneous cluster topology, $N$ for the number of ranks in the current cluster, $B$ for the number of border ranks, and $n$ for the data count of the send buffer of the collective operation.
\caption{\small Cluster-level Primitive Breakdown of Heterogeneous Collectives}
\label{tab:collectives-breakdown}
\end{table*}

%% file: fig_design-c2c-algo-ag.tex
\begin{figure}[!th]
    \centering
    \includegraphics[width=.75\textwidth]{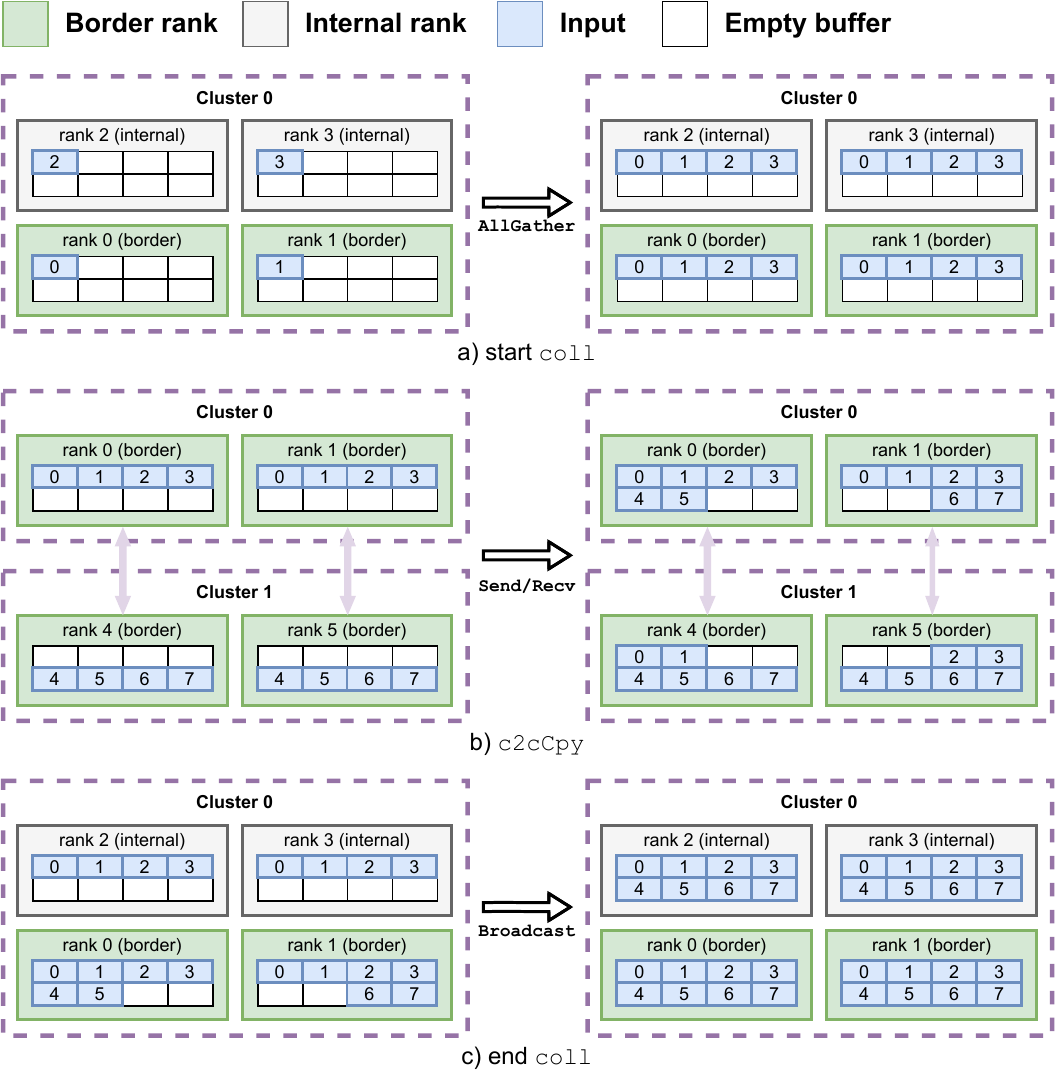}
    \caption{\small Collective breakdown of \texttt{AllGather}}
    \label{fig:design-c2c-ag}
\end{figure}

%% file: fig_design-c2c-algo-ar.tex
\begin{figure}[!hb]
    \centering
    \includegraphics[width=.75\textwidth]{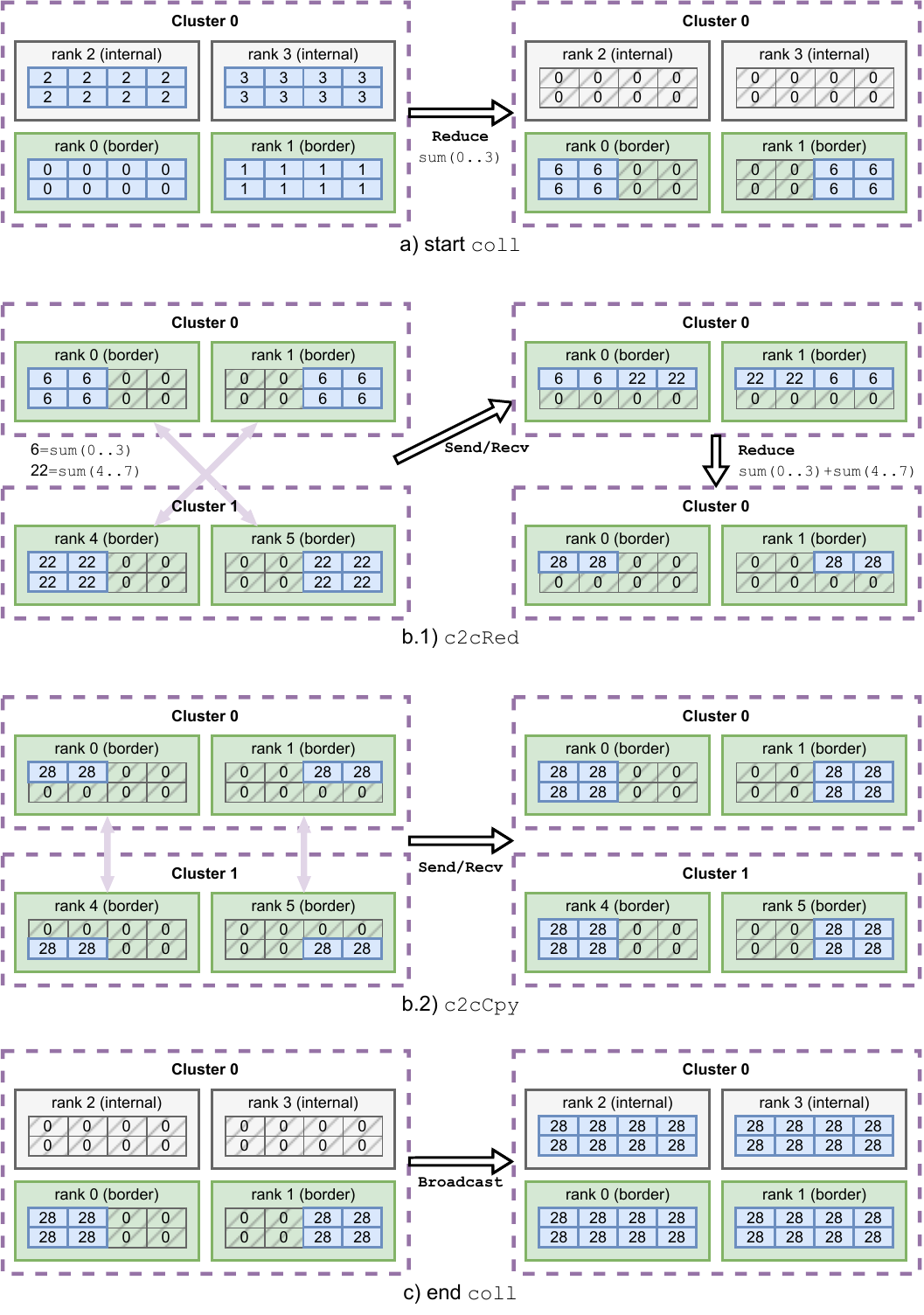}
    \caption{\small Collective breakdown of \texttt{AllReduce}}
    \label{fig:design-c2c-ar}
\end{figure}

%% file: tab_eval-e2e-setup.tex
\begin{table*}[!h]
\centering
\resizebox{.75\textwidth}{!}{
    \begin{tabular}{|c|c|c|} \hline
         & \textbf{Setup 1} & \textbf{Setup 2} \\ \hline
        \textbf{LLM} & Llama3-3B & Llama3-8B \\ \hline
        \textbf{\#Layers} & 36 & 32 \\ \hline
        \textbf{Global Batchsize} & 128 & 256 \\ \hline
        \textbf{Server Model} & 1 $\times$ A800 + 1 $\times$ HW-1 & 2 $\times$ A800 + 2 $\times$ HW-1 \\ \hline
        \textbf{Accelerators} & NV:$8$,V1:$16$ & NV:$8 \times 2$,V1:$16 \times 2$ \\ \hline
        \textbf{Parallel Strategy} & PP: (NV:$26$,V1:$10$), DP: (NV:$8$,V1:$16$) & PP: (NV:$10,11$,V1:$6,5$), DP: (NV:$8$,V1:$16$) \\ \hline
        \textbf{CCL} & \multicolumn{2}{c|}{Gloo / \sysname} \\ \hline
    \end{tabular}
}
    \caption{\small End-to-end training performance evaluation setup}
    \label{tab:eval-e2e-setup-perf}
\end{table*}

\begin{table*}[!h]
\centering
\resizebox{\textwidth}{!}{
    \begin{tabular}{
    |>{\centering\arraybackslash}m{3cm}
    |>{\centering\arraybackslash}m{3cm}
    |>{\centering\arraybackslash}m{3cm}
    |>{\centering\arraybackslash}m{3cm}
    |>{\centering\arraybackslash}m{3cm}
    |>{\centering\arraybackslash}m{3cm}
    |} \hline
         & \textbf{Setup 3} & \textbf{Setup 4} & \textbf{Setup 5} & \textbf{Setup 6} & \textbf{Setup 7} \\ \hline
        \textbf{LLM} & \multicolumn{5}{c|}{Llama3-8B} \\ \hline
        \textbf{\#Layers} & \multicolumn{5}{c|}{32} \\ \hline
        \textbf{Global Batchsize} & \multicolumn{5}{c|}{512} \\ \hline
        \textbf{Server Model} & 2 $\times$ A800 & 2 $\times$ HW-3 & 1 $\times$ A800 + 1 $\times$ HW-3 & 2 $\times$ A800 + 2 $\times$ HW-3 & 4 $\times$ A800 + 4 $\times$ HW-3 \\ \hline
        \textbf{Accelerators} & NV:$8 \times 2$ & V3:$8 \times 2$ & NV:$8$,V3:$8$ & NV:$8 \times 2$,V3:$8 \times 2$ & NV:$8 \times 4$,V3:$8 \times 4$ \\ \hline
        \multirow{3}{*}{\textbf{Parallel Strategy}} & TP:2,DP:8,PP:1 & TP:2,DP:8,PP:1 & TP:2,DP:8,PP:1 & TP:2,DP:16,PP:1 & - \\ \cline{2-6}
        & TP:1,DP:8,PP:2 & TP:1,DP:8,PP:2 & TP:1,DP:8,PP:2 & TP:1,DP:16,PP:2 & TP:1,DP:32,PP:2 \\ \cline{2-6}
        & - & - & - & TP:2,DP:8,PP:2 & TP:2,DP:16,PP:2 \\ \hline
        \textbf{CCL} & NCCL & \cccl & \multicolumn{3}{c|}{\sysname} \\ \hline
    \end{tabular}
}
    \caption{\small End-to-end training scalability evaluation setup}
    \label{tab:eval-e2e-setup-scale}
\end{table*}

%% file: tab_eval-serving-setup.tex
\begin{table}[!h]
\centering
\resizebox{.4\textwidth}{!}{
    \begin{tabular}{|c|c|c|} \hline
         & \textbf{Setup 1} & \textbf{Setup 2} \\ \hline
        \textbf{LLM} & Qwen2-7B & Qwen2-7B \\ \hline
        \textbf{Server} & 2 $\times$ A800 & 1 $\times$ A800 + 1 $\times$ HW-3 \\ \hline
        \textbf{Accelerators} & NV:2 & NV:1,V3:1 \\ \hline
        \textbf{Prefill} & NV:1 & NV:1 \\ \hline
        \textbf{Decode} & NV:1 & V3:1 \\ \hline
        \multirow{3}{*}{\textbf{CCL}} & NCCL & - \\ \cline{2-3}
         & Host & Host \\ \cline{2-3}
         & \sysname & \sysname \\ \hline
    \end{tabular}
}
    \caption{\small LLM serving evaluation setup.}
    \label{tab:eval-serving-setup}
\end{table}

%% file: main.bbl

\begin{thebibliography}{47}


\ifx \showCODEN    \undefined \def \showCODEN     #1{\unskip}     \fi
\ifx \showDOI      \undefined \def \showDOI       #1{#1}\fi
\ifx \showISBNx    \undefined \def \showISBNx     #1{\unskip}     \fi
\ifx \showISBNxiii \undefined \def \showISBNxiii  #1{\unskip}     \fi
\ifx \showISSN     \undefined \def \showISSN      #1{\unskip}     \fi
\ifx \showLCCN     \undefined \def \showLCCN      #1{\unskip}     \fi
\ifx \shownote     \undefined \def \shownote      #1{#1}          \fi
\ifx \showarticletitle \undefined \def \showarticletitle #1{#1}   \fi
\ifx \showURL      \undefined \def \showURL       {\relax}        \fi
\providecommand\bibfield[2]{#2}
\providecommand\bibinfo[2]{#2}
\providecommand\natexlab[1]{#1}
\providecommand\showeprint[2][]{arXiv:#2}

\bibitem[\protect\citeauthoryear{??}{ope}{2024}]%
        {openmpi}
 \bibinfo{year}{2024}\natexlab{}.
\newblock \bibinfo{title}{OpenMPI}.
\newblock   (\bibinfo{year}{2024}).
\newblock
\newblock
\shownote{\url{https://www.open-mpi.org/}.}


\bibitem[\protect\citeauthoryear{AMD}{AMD}{2024a}]%
        {mi300x}
\bibfield{author}{\bibinfo{person}{AMD}.} \bibinfo{year}{2024}\natexlab{a}.
\newblock \bibinfo{title}{AMD Instinct Accelerators}.
\newblock   (\bibinfo{year}{2024}).
\newblock
\newblock
\shownote{\url{https://www.amd.com/en/products/accelerators/instinct.html}.}


\bibitem[\protect\citeauthoryear{AMD}{AMD}{2024b}]%
        {rccl}
\bibfield{author}{\bibinfo{person}{AMD}.} \bibinfo{year}{2024}\natexlab{b}.
\newblock \bibinfo{title}{RCCL}.
\newblock   (\bibinfo{year}{2024}).
\newblock
\newblock
\shownote{\url{https://github.com/ROCm/rccl}.}


\bibitem[\protect\citeauthoryear{AMD}{AMD}{2024c}]%
        {rocm}
\bibfield{author}{\bibinfo{person}{AMD}.} \bibinfo{year}{2024}\natexlab{c}.
\newblock \bibinfo{title}{ROCm}.
\newblock   (\bibinfo{year}{2024}).
\newblock
\newblock
\shownote{\url{https://www.amd.com/en/products/software/rocm.html}.}


\bibitem[\protect\citeauthoryear{An, Bi, Chen, Chen, Deng, Ding, Dong, Du, Gao, Guan, et~al\mbox{.}}{An et~al\mbox{.}}{2024}]%
        {an2024fire}
\bibfield{author}{\bibinfo{person}{Wei An}, \bibinfo{person}{Xiao Bi}, \bibinfo{person}{Guanting Chen}, \bibinfo{person}{Shanhuang Chen}, \bibinfo{person}{Chengqi Deng}, \bibinfo{person}{Honghui Ding}, \bibinfo{person}{Kai Dong}, \bibinfo{person}{Qiushi Du}, \bibinfo{person}{Wenjun Gao}, \bibinfo{person}{Kang Guan}, {et~al\mbox{.}}} \bibinfo{year}{2024}\natexlab{}.
\newblock \showarticletitle{Fire-Flyer AI-HPC: A Cost-Effective Software-Hardware Co-Design for Deep Learning}. In \bibinfo{booktitle}{{\em SC24: International Conference for High Performance Computing, Networking, Storage and Analysis}}. IEEE, \bibinfo{pages}{1--23}.
\newblock


\bibitem[\protect\citeauthoryear{Cai, Liu, Maleki, Musuvathi, Mytkowicz, Nelson, and Saarikivi}{Cai et~al\mbox{.}}{2021}]%
        {cai2021synthesizing}
\bibfield{author}{\bibinfo{person}{Zixian Cai}, \bibinfo{person}{Zhengyang Liu}, \bibinfo{person}{Saeed Maleki}, \bibinfo{person}{Madanlal Musuvathi}, \bibinfo{person}{Todd Mytkowicz}, \bibinfo{person}{Jacob Nelson}, {and} \bibinfo{person}{Olli Saarikivi}.} \bibinfo{year}{2021}\natexlab{}.
\newblock \showarticletitle{Synthesizing optimal collective algorithms}. In \bibinfo{booktitle}{{\em Proceedings of the 26th ACM SIGPLAN Symposium on Principles and Practice of Parallel Programming}}. \bibinfo{pages}{62--75}.
\newblock


\bibitem[\protect\citeauthoryear{Chen, Shafie~Khorassani, Kousha, Zhou, Yao, Subramoni, and Panda}{Chen et~al\mbox{.}}{2023}]%
        {chen2023mpi}
\bibfield{author}{\bibinfo{person}{Chen-Chun Chen}, \bibinfo{person}{Kawthar Shafie~Khorassani}, \bibinfo{person}{Pouya Kousha}, \bibinfo{person}{Qinghua Zhou}, \bibinfo{person}{Jinghan Yao}, \bibinfo{person}{Hari Subramoni}, {and} \bibinfo{person}{Dhabaleswar~K Panda}.} \bibinfo{year}{2023}\natexlab{}.
\newblock \showarticletitle{MPI-xCCL: A Portable MPI Library over Collective Communication Libraries for Various Accelerators}. In \bibinfo{booktitle}{{\em Proceedings of the SC'23 Workshops of The International Conference on High Performance Computing, Network, Storage, and Analysis}}. \bibinfo{pages}{847--854}.
\newblock


\bibitem[\protect\citeauthoryear{Chen, Zhang, Du, Xiang, Yue, Zhang, Cai, and Zhang}{Chen et~al\mbox{.}}{2024}]%
        {chen2024understanding}
\bibfield{author}{\bibinfo{person}{Hongzheng Chen}, \bibinfo{person}{Jiahao Zhang}, \bibinfo{person}{Yixiao Du}, \bibinfo{person}{Shaojie Xiang}, \bibinfo{person}{Zichao Yue}, \bibinfo{person}{Niansong Zhang}, \bibinfo{person}{Yaohui Cai}, {and} \bibinfo{person}{Zhiru Zhang}.} \bibinfo{year}{2024}\natexlab{}.
\newblock \showarticletitle{Understanding the potential of fpga-based spatial acceleration for large language model inference}.
\newblock \bibinfo{journal}{{\em ACM Transactions on Reconfigurable Technology and Systems\/}} (\bibinfo{year}{2024}).
\newblock


\bibitem[\protect\citeauthoryear{Consortium}{Consortium}{2024a}]%
        {openucx-git}
\bibfield{author}{\bibinfo{person}{UCF Consortium}.} \bibinfo{year}{2024}\natexlab{a}.
\newblock \bibinfo{title}{Unified Communication X Library Source Code \url{https://github.com/openucx/ucx}}.
\newblock   (\bibinfo{year}{2024}).
\newblock


\bibitem[\protect\citeauthoryear{Consortium}{Consortium}{2024b}]%
        {openucx-home}
\bibfield{author}{\bibinfo{person}{UCF Consortium}.} \bibinfo{year}{2024}\natexlab{b}.
\newblock \bibinfo{title}{Unified Communication X \url{https://openucx.org/}}.
\newblock   (\bibinfo{year}{2024}).
\newblock


\bibitem[\protect\citeauthoryear{Corporation}{Corporation}{2026a}]%
        {nvidia_gdr}
\bibfield{author}{\bibinfo{person}{NVIDIA Corporation}.} \bibinfo{year}{2026}\natexlab{a}.
\newblock \bibinfo{title}{GPU-Direct RDMA (GDR)}.
\newblock \bibinfo{howpublished}{\url{https://developer.nvidia.com/gpudirect}}.   (\bibinfo{year}{2026}).
\newblock
\newblock
\shownote{Accessed: 2026-02-07.}


\bibitem[\protect\citeauthoryear{Corporation}{Corporation}{2026b}]%
        {nvidia_nvshmem}
\bibfield{author}{\bibinfo{person}{NVIDIA Corporation}.} \bibinfo{year}{2026}\natexlab{b}.
\newblock \bibinfo{title}{NVSHMEM: NVIDIA SHMEM Library}.
\newblock \bibinfo{howpublished}{\url{https://developer.nvidia.com/nvshmem}}.   (\bibinfo{year}{2026}).
\newblock
\newblock
\shownote{Accessed: 2026-02-07.}


\bibitem[\protect\citeauthoryear{DeepSeek-AI, Liu, Feng, Xue, Wang, Wu, Lu, Zhao, Deng, Zhang, Ruan, Dai, Guo, Yang, Chen, Ji, Li, Lin, Dai, Luo, Hao, Chen, Li, Zhang, Bao, Xu, Wang, Zhang, Ding, Xin, Gao, Li, Qu, Cai, Liang, Guo, Ni, Li, Wang, Chen, Chen, Yuan, Qiu, Li, Song, Dong, Hu, Gao, Guan, Huang, Yu, Wang, Zhang, Xu, Xia, Zhao, Wang, Zhang, Li, Wang, Zhang, Zhang, Tang, Li, Tian, Huang, Wang, Zhang, Wang, Zhu, Chen, Du, Chen, Jin, Ge, Zhang, Pan, Wang, Xu, Zhang, Chen, Li, Lu, Zhou, Chen, Wu, Ye, Ye, Ma, Wang, Zhou, Yu, Zhou, Pan, Wang, Yun, Pei, Sun, Xiao, Zeng, Zhao, An, Liu, Liang, Gao, Yu, Zhang, Li, Jin, Wang, Bi, Liu, Wang, Shen, Chen, Zhang, Chen, Nie, Sun, Wang, Cheng, Liu, Xie, Liu, Yu, Song, Shan, Zhou, Yang, Li, Su, Lin, Li, Wang, Wei, Zhu, Zhang, Xu, Xu, Huang, Li, Zhao, Sun, Li, Wang, Yu, Zheng, Zhang, Shi, Xiong, He, Tang, Piao, Wang, Tan, Ma, Liu, Guo, Wu, Ou, Zhu, Wang, Gong, Zou, He, Zha, Xiong, Ma, Yan, Luo, You, Liu, Zhou, Wu, Ren, Ren, Sha, Fu, Xu, Huang, Zhang, Xie, Zhang, Hao,
  Gou, Ma, Yan, Shao, Xu, Wu, Zhang, Li, Gu, Zhu, Liu, Li, Xie, Song, Gao, and Pan}{DeepSeek-AI et~al\mbox{.}}{2024}]%
        {deepseekv3}
\bibfield{author}{\bibinfo{person}{DeepSeek-AI}, \bibinfo{person}{Aixin Liu}, \bibinfo{person}{Bei Feng}, \bibinfo{person}{Bing Xue}, \bibinfo{person}{Bingxuan Wang}, \bibinfo{person}{Bochao Wu}, \bibinfo{person}{Chengda Lu}, \bibinfo{person}{Chenggang Zhao}, \bibinfo{person}{Chengqi Deng}, \bibinfo{person}{Chenyu Zhang}, \bibinfo{person}{Chong Ruan}, \bibinfo{person}{Damai Dai}, \bibinfo{person}{Daya Guo}, \bibinfo{person}{Dejian Yang}, \bibinfo{person}{Deli Chen}, \bibinfo{person}{Dongjie Ji}, \bibinfo{person}{Erhang Li}, \bibinfo{person}{Fangyun Lin}, \bibinfo{person}{Fucong Dai}, \bibinfo{person}{Fuli Luo}, \bibinfo{person}{Guangbo Hao}, \bibinfo{person}{Guanting Chen}, \bibinfo{person}{Guowei Li}, \bibinfo{person}{H. Zhang}, \bibinfo{person}{Han Bao}, \bibinfo{person}{Hanwei Xu}, \bibinfo{person}{Haocheng Wang}, \bibinfo{person}{Haowei Zhang}, \bibinfo{person}{Honghui Ding}, \bibinfo{person}{Huajian Xin}, \bibinfo{person}{Huazuo Gao}, \bibinfo{person}{Hui Li}, \bibinfo{person}{Hui Qu},
  \bibinfo{person}{J.~L. Cai}, \bibinfo{person}{Jian Liang}, \bibinfo{person}{Jianzhong Guo}, \bibinfo{person}{Jiaqi Ni}, \bibinfo{person}{Jiashi Li}, \bibinfo{person}{Jiawei Wang}, \bibinfo{person}{Jin Chen}, \bibinfo{person}{Jingchang Chen}, \bibinfo{person}{Jingyang Yuan}, \bibinfo{person}{Junjie Qiu}, \bibinfo{person}{Junlong Li}, \bibinfo{person}{Junxiao Song}, \bibinfo{person}{Kai Dong}, \bibinfo{person}{Kai Hu}, \bibinfo{person}{Kaige Gao}, \bibinfo{person}{Kang Guan}, \bibinfo{person}{Kexin Huang}, \bibinfo{person}{Kuai Yu}, \bibinfo{person}{Lean Wang}, \bibinfo{person}{Lecong Zhang}, \bibinfo{person}{Lei Xu}, \bibinfo{person}{Leyi Xia}, \bibinfo{person}{Liang Zhao}, \bibinfo{person}{Litong Wang}, \bibinfo{person}{Liyue Zhang}, \bibinfo{person}{Meng Li}, \bibinfo{person}{Miaojun Wang}, \bibinfo{person}{Mingchuan Zhang}, \bibinfo{person}{Minghua Zhang}, \bibinfo{person}{Minghui Tang}, \bibinfo{person}{Mingming Li}, \bibinfo{person}{Ning Tian}, \bibinfo{person}{Panpan Huang}, \bibinfo{person}{Peiyi
  Wang}, \bibinfo{person}{Peng Zhang}, \bibinfo{person}{Qiancheng Wang}, \bibinfo{person}{Qihao Zhu}, \bibinfo{person}{Qinyu Chen}, \bibinfo{person}{Qiushi Du}, \bibinfo{person}{R.~J. Chen}, \bibinfo{person}{R.~L. Jin}, \bibinfo{person}{Ruiqi Ge}, \bibinfo{person}{Ruisong Zhang}, \bibinfo{person}{Ruizhe Pan}, \bibinfo{person}{Runji Wang}, \bibinfo{person}{Runxin Xu}, \bibinfo{person}{Ruoyu Zhang}, \bibinfo{person}{Ruyi Chen}, \bibinfo{person}{S.~S. Li}, \bibinfo{person}{Shanghao Lu}, \bibinfo{person}{Shangyan Zhou}, \bibinfo{person}{Shanhuang Chen}, \bibinfo{person}{Shaoqing Wu}, \bibinfo{person}{Shengfeng Ye}, \bibinfo{person}{Shengfeng Ye}, \bibinfo{person}{Shirong Ma}, \bibinfo{person}{Shiyu Wang}, \bibinfo{person}{Shuang Zhou}, \bibinfo{person}{Shuiping Yu}, \bibinfo{person}{Shunfeng Zhou}, \bibinfo{person}{Shuting Pan}, \bibinfo{person}{T. Wang}, \bibinfo{person}{Tao Yun}, \bibinfo{person}{Tian Pei}, \bibinfo{person}{Tianyu Sun}, \bibinfo{person}{W.~L. Xiao}, \bibinfo{person}{Wangding Zeng},
  \bibinfo{person}{Wanjia Zhao}, \bibinfo{person}{Wei An}, \bibinfo{person}{Wen Liu}, \bibinfo{person}{Wenfeng Liang}, \bibinfo{person}{Wenjun Gao}, \bibinfo{person}{Wenqin Yu}, \bibinfo{person}{Wentao Zhang}, \bibinfo{person}{X.~Q. Li}, \bibinfo{person}{Xiangyue Jin}, \bibinfo{person}{Xianzu Wang}, \bibinfo{person}{Xiao Bi}, \bibinfo{person}{Xiaodong Liu}, \bibinfo{person}{Xiaohan Wang}, \bibinfo{person}{Xiaojin Shen}, \bibinfo{person}{Xiaokang Chen}, \bibinfo{person}{Xiaokang Zhang}, \bibinfo{person}{Xiaosha Chen}, \bibinfo{person}{Xiaotao Nie}, \bibinfo{person}{Xiaowen Sun}, \bibinfo{person}{Xiaoxiang Wang}, \bibinfo{person}{Xin Cheng}, \bibinfo{person}{Xin Liu}, \bibinfo{person}{Xin Xie}, \bibinfo{person}{Xingchao Liu}, \bibinfo{person}{Xingkai Yu}, \bibinfo{person}{Xinnan Song}, \bibinfo{person}{Xinxia Shan}, \bibinfo{person}{Xinyi Zhou}, \bibinfo{person}{Xinyu Yang}, \bibinfo{person}{Xinyuan Li}, \bibinfo{person}{Xuecheng Su}, \bibinfo{person}{Xuheng Lin}, \bibinfo{person}{Y.~K. Li},
  \bibinfo{person}{Y.~Q. Wang}, \bibinfo{person}{Y.~X. Wei}, \bibinfo{person}{Y.~X. Zhu}, \bibinfo{person}{Yang Zhang}, \bibinfo{person}{Yanhong Xu}, \bibinfo{person}{Yanhong Xu}, \bibinfo{person}{Yanping Huang}, \bibinfo{person}{Yao Li}, \bibinfo{person}{Yao Zhao}, \bibinfo{person}{Yaofeng Sun}, \bibinfo{person}{Yaohui Li}, \bibinfo{person}{Yaohui Wang}, \bibinfo{person}{Yi Yu}, \bibinfo{person}{Yi Zheng}, \bibinfo{person}{Yichao Zhang}, \bibinfo{person}{Yifan Shi}, \bibinfo{person}{Yiliang Xiong}, \bibinfo{person}{Ying He}, \bibinfo{person}{Ying Tang}, \bibinfo{person}{Yishi Piao}, \bibinfo{person}{Yisong Wang}, \bibinfo{person}{Yixuan Tan}, \bibinfo{person}{Yiyang Ma}, \bibinfo{person}{Yiyuan Liu}, \bibinfo{person}{Yongqiang Guo}, \bibinfo{person}{Yu Wu}, \bibinfo{person}{Yuan Ou}, \bibinfo{person}{Yuchen Zhu}, \bibinfo{person}{Yuduan Wang}, \bibinfo{person}{Yue Gong}, \bibinfo{person}{Yuheng Zou}, \bibinfo{person}{Yujia He}, \bibinfo{person}{Yukun Zha}, \bibinfo{person}{Yunfan Xiong},
  \bibinfo{person}{Yunxian Ma}, \bibinfo{person}{Yuting Yan}, \bibinfo{person}{Yuxiang Luo}, \bibinfo{person}{Yuxiang You}, \bibinfo{person}{Yuxuan Liu}, \bibinfo{person}{Yuyang Zhou}, \bibinfo{person}{Z.~F. Wu}, \bibinfo{person}{Z.~Z. Ren}, \bibinfo{person}{Zehui Ren}, \bibinfo{person}{Zhangli Sha}, \bibinfo{person}{Zhe Fu}, \bibinfo{person}{Zhean Xu}, \bibinfo{person}{Zhen Huang}, \bibinfo{person}{Zhen Zhang}, \bibinfo{person}{Zhenda Xie}, \bibinfo{person}{Zhengyan Zhang}, \bibinfo{person}{Zhewen Hao}, \bibinfo{person}{Zhibin Gou}, \bibinfo{person}{Zhicheng Ma}, \bibinfo{person}{Zhigang Yan}, \bibinfo{person}{Zhihong Shao}, \bibinfo{person}{Zhipeng Xu}, \bibinfo{person}{Zhiyu Wu}, \bibinfo{person}{Zhongyu Zhang}, \bibinfo{person}{Zhuoshu Li}, \bibinfo{person}{Zihui Gu}, \bibinfo{person}{Zijia Zhu}, \bibinfo{person}{Zijun Liu}, \bibinfo{person}{Zilin Li}, \bibinfo{person}{Ziwei Xie}, \bibinfo{person}{Ziyang Song}, \bibinfo{person}{Ziyi Gao}, {and} \bibinfo{person}{Zizheng Pan}.}
  \bibinfo{year}{2024}\natexlab{}.
\newblock \bibinfo{title}{DeepSeek-V3 Technical Report}.
\newblock   (\bibinfo{year}{2024}).
\newblock
\showeprint[arxiv]{cs.CL/2412.19437}
\showURL{%
\url{https://arxiv.org/abs/2412.19437}}


\bibitem[\protect\citeauthoryear{Facebook}{Facebook}{2024}]%
        {gloo}
\bibfield{author}{\bibinfo{person}{Facebook}.} \bibinfo{year}{2024}\natexlab{}.
\newblock \bibinfo{title}{Gloo}.
\newblock   (\bibinfo{year}{2024}).
\newblock
\newblock
\shownote{\url{https://github.com/facebookincubator/gloo/}.}


\bibitem[\protect\citeauthoryear{Fedus, Zoph, and Shazeer}{Fedus et~al\mbox{.}}{2022}]%
        {switch_transformer}
\bibfield{author}{\bibinfo{person}{William Fedus}, \bibinfo{person}{Barret Zoph}, {and} \bibinfo{person}{Noam Shazeer}.} \bibinfo{year}{2022}\natexlab{}.
\newblock \showarticletitle{Switch transformers: Scaling to trillion parameter models with simple and efficient sparsity}.
\newblock \bibinfo{journal}{{\em Journal of Machine Learning Research\/}} \bibinfo{volume}{23}, \bibinfo{number}{120} (\bibinfo{year}{2022}), \bibinfo{pages}{1--39}.
\newblock


\bibitem[\protect\citeauthoryear{Graphcore}{Graphcore}{2024}]%
        {graphcore}
\bibfield{author}{\bibinfo{person}{Graphcore}.} \bibinfo{year}{2024}\natexlab{}.
\newblock \bibinfo{title}{Graphcore}.
\newblock   (\bibinfo{year}{2024}).
\newblock
\newblock
\shownote{\url{https://www.graphcore.ai/}.}


\bibitem[\protect\citeauthoryear{Hamidouche, Venkatesh, Awan, Subramoni, Chu, and Panda}{Hamidouche et~al\mbox{.}}{2015}]%
        {hamidouche2015exploiting}
\bibfield{author}{\bibinfo{person}{Khaled Hamidouche}, \bibinfo{person}{Akshay Venkatesh}, \bibinfo{person}{Ammar~Ahmad Awan}, \bibinfo{person}{Hari Subramoni}, \bibinfo{person}{Ching-Hsiang Chu}, {and} \bibinfo{person}{Dhabaleswar~K Panda}.} \bibinfo{year}{2015}\natexlab{}.
\newblock \showarticletitle{Exploiting GPUDirect RDMA in designing high performance OpenSHMEM for NVIDIA GPU clusters}. In \bibinfo{booktitle}{{\em 2015 IEEE International Conference on Cluster Computing}}. IEEE, \bibinfo{pages}{78--87}.
\newblock


\bibitem[\protect\citeauthoryear{Hong, Moon, Kim, Lee, Kim, Lee, and Kim}{Hong et~al\mbox{.}}{2022}]%
        {hong2022dfx}
\bibfield{author}{\bibinfo{person}{Seongmin Hong}, \bibinfo{person}{Seungjae Moon}, \bibinfo{person}{Junsoo Kim}, \bibinfo{person}{Sungjae Lee}, \bibinfo{person}{Minsub Kim}, \bibinfo{person}{Dongsoo Lee}, {and} \bibinfo{person}{Joo-Young Kim}.} \bibinfo{year}{2022}\natexlab{}.
\newblock \showarticletitle{Dfx: A low-latency multi-fpga appliance for accelerating transformer-based text generation}. In \bibinfo{booktitle}{{\em 2022 55th IEEE/ACM International Symposium on Microarchitecture (MICRO)}}. IEEE, \bibinfo{pages}{616--630}.
\newblock


\bibitem[\protect\citeauthoryear{Huang, Wan, Ye, Jha, Wang, Li, Zhang, and Chen}{Huang et~al\mbox{.}}{2024}]%
        {huang2024new}
\bibfield{author}{\bibinfo{person}{Yingbing Huang}, \bibinfo{person}{Lily~Jiaxin Wan}, \bibinfo{person}{Hanchen Ye}, \bibinfo{person}{Manvi Jha}, \bibinfo{person}{Jinghua Wang}, \bibinfo{person}{Yuhong Li}, \bibinfo{person}{Xiaofan Zhang}, {and} \bibinfo{person}{Deming Chen}.} \bibinfo{year}{2024}\natexlab{}.
\newblock \showarticletitle{New solutions on LLM acceleration, optimization, and application}. In \bibinfo{booktitle}{{\em Proceedings of the 61st ACM/IEEE Design Automation Conference}}. \bibinfo{pages}{1--4}.
\newblock


\bibitem[\protect\citeauthoryear{Huawei}{Huawei}{2024}]%
        {ascend}
\bibfield{author}{\bibinfo{person}{Huawei}.} \bibinfo{year}{2024}\natexlab{}.
\newblock \bibinfo{title}{Ascend Computing}.
\newblock   (\bibinfo{year}{2024}).
\newblock
\newblock
\shownote{\url{https://e.huawei.com/en/products/computing/ascend}.}


\bibitem[\protect\citeauthoryear{Intel}{Intel}{2024}]%
        {oneccl}
\bibfield{author}{\bibinfo{person}{Intel}.} \bibinfo{year}{2024}\natexlab{}.
\newblock \bibinfo{title}{OneCCL}.
\newblock   (\bibinfo{year}{2024}).
\newblock
\newblock
\shownote{\url{https://www.intel.com/content/www/us/en/developer/tools/oneapi/oneccl.html}.}


\bibitem[\protect\citeauthoryear{Jia, Jiang, Wang, Xiao, Shi, Zhang, Li, Chen, Li, Zheng, et~al\mbox{.}}{Jia et~al\mbox{.}}{2022}]%
        {jia2022whale}
\bibfield{author}{\bibinfo{person}{Xianyan Jia}, \bibinfo{person}{Le Jiang}, \bibinfo{person}{Ang Wang}, \bibinfo{person}{Wencong Xiao}, \bibinfo{person}{Ziji Shi}, \bibinfo{person}{Jie Zhang}, \bibinfo{person}{Xinyuan Li}, \bibinfo{person}{Langshi Chen}, \bibinfo{person}{Yong Li}, \bibinfo{person}{Zhen Zheng}, {et~al\mbox{.}}} \bibinfo{year}{2022}\natexlab{}.
\newblock \showarticletitle{Whale: Efficient giant model training over heterogeneous $\{$GPUs$\}$}. In \bibinfo{booktitle}{{\em 2022 USENIX Annual Technical Conference (USENIX ATC 22)}}. \bibinfo{pages}{673--688}.
\newblock


\bibitem[\protect\citeauthoryear{Kachris}{Kachris}{2025}]%
        {kachris2025survey}
\bibfield{author}{\bibinfo{person}{Christoforos Kachris}.} \bibinfo{year}{2025}\natexlab{}.
\newblock \showarticletitle{A survey on hardware accelerators for large language models}.
\newblock \bibinfo{journal}{{\em Applied Sciences\/}} \bibinfo{volume}{15}, \bibinfo{number}{2} (\bibinfo{year}{2025}), \bibinfo{pages}{586}.
\newblock


\bibitem[\protect\citeauthoryear{Khronos}{Khronos}{2024}]%
        {sycl}
\bibfield{author}{\bibinfo{person}{Khronos}.} \bibinfo{year}{2024}\natexlab{}.
\newblock \bibinfo{title}{SYCL}.
\newblock   (\bibinfo{year}{2024}).
\newblock
\newblock
\shownote{\url{https://www.khronos.org/sycl/}.}


\bibitem[\protect\citeauthoryear{Kim, Ryu, and Lee}{Kim et~al\mbox{.}}{2024}]%
        {kim2024tccl}
\bibfield{author}{\bibinfo{person}{Heehoon Kim}, \bibinfo{person}{Junyeol Ryu}, {and} \bibinfo{person}{Jaejin Lee}.} \bibinfo{year}{2024}\natexlab{}.
\newblock \showarticletitle{TCCL: Discovering Better Communication Paths for PCIe GPU Clusters}. In \bibinfo{booktitle}{{\em Proceedings of the 29th ACM International Conference on Architectural Support for Programming Languages and Operating Systems, Volume 3}}. \bibinfo{pages}{999--1015}.
\newblock


\bibitem[\protect\citeauthoryear{Liu, Arzani, Kakarla, Zhao, Liu, Castro, Kandula, and Marshall}{Liu et~al\mbox{.}}{2024}]%
        {liu2024rethinking}
\bibfield{author}{\bibinfo{person}{Xuting Liu}, \bibinfo{person}{Behnaz Arzani}, \bibinfo{person}{Siva Kesava~Reddy Kakarla}, \bibinfo{person}{Liangyu Zhao}, \bibinfo{person}{Vincent Liu}, \bibinfo{person}{Miguel Castro}, \bibinfo{person}{Srikanth Kandula}, {and} \bibinfo{person}{Luke Marshall}.} \bibinfo{year}{2024}\natexlab{}.
\newblock \showarticletitle{Rethinking machine learning collective communication as a multi-commodity flow problem}. In \bibinfo{booktitle}{{\em Proceedings of the ACM SIGCOMM 2024 Conference}}. \bibinfo{pages}{16--37}.
\newblock


\bibitem[\protect\citeauthoryear{{Llama Team, AI @ Meta}}{{Llama Team, AI @ Meta}}{2024}]%
        {llama3herdmodels}
\bibfield{author}{\bibinfo{person}{{Llama Team, AI @ Meta}}.} \bibinfo{year}{2024}\natexlab{}.
\newblock \bibinfo{title}{The Llama 3 Herd of Models}.
\newblock   (\bibinfo{year}{2024}).
\newblock
\showeprint[arxiv]{cs.AI/2407.21783}
\showURL{%
\url{https://arxiv.org/abs/2407.21783}}


\bibitem[\protect\citeauthoryear{Microsoft}{Microsoft}{2024}]%
        {msccl}
\bibfield{author}{\bibinfo{person}{Microsoft}.} \bibinfo{year}{2024}\natexlab{}.
\newblock \bibinfo{title}{MSCCL}.
\newblock   (\bibinfo{year}{2024}).
\newblock
\newblock
\shownote{\url{https://github.com/microsoft/msccl}.}


\bibitem[\protect\citeauthoryear{NVIDIA}{NVIDIA}{2024a}]%
        {a100}
\bibfield{author}{\bibinfo{person}{NVIDIA}.} \bibinfo{year}{2024}\natexlab{a}.
\newblock \bibinfo{title}{A100}.
\newblock   (\bibinfo{year}{2024}).
\newblock
\newblock
\shownote{\url{https://www.nvidia.com/en-us/data-center/a100/}.}


\bibitem[\protect\citeauthoryear{NVIDIA}{NVIDIA}{2024b}]%
        {nccl}
\bibfield{author}{\bibinfo{person}{NVIDIA}.} \bibinfo{year}{2024}\natexlab{b}.
\newblock \bibinfo{title}{NCCL}.
\newblock   (\bibinfo{year}{2024}).
\newblock
\newblock
\shownote{\url{https://developer.nvidia.com/nccl}.}


\bibitem[\protect\citeauthoryear{Organization}{Organization}{2024}]%
        {openacc}
\bibfield{author}{\bibinfo{person}{OpenACC Organization}.} \bibinfo{year}{2024}\natexlab{}.
\newblock \bibinfo{title}{OpenACC}.
\newblock   (\bibinfo{year}{2024}).
\newblock
\newblock
\shownote{\url{https://www.openacc.org/}.}


\bibitem[\protect\citeauthoryear{Park, Yun, Chang, Nguyen, Lee, Choi, Noh, and Choi}{Park et~al\mbox{.}}{2020}]%
        {park2020hetpipe}
\bibfield{author}{\bibinfo{person}{Jay~H Park}, \bibinfo{person}{Gyeongchan Yun}, \bibinfo{person}{M~Yi Chang}, \bibinfo{person}{Nguyen~T Nguyen}, \bibinfo{person}{Seungmin Lee}, \bibinfo{person}{Jaesik Choi}, \bibinfo{person}{Sam~H Noh}, {and} \bibinfo{person}{Young-ri Choi}.} \bibinfo{year}{2020}\natexlab{}.
\newblock \showarticletitle{$\{$HetPipe$\}$: Enabling large $\{$DNN$\}$ training on (whimpy) heterogeneous $\{$GPU$\}$ clusters through integration of pipelined model parallelism and data parallelism}. In \bibinfo{booktitle}{{\em 2020 USENIX Annual Technical Conference (USENIX ATC 20)}}. \bibinfo{pages}{307--321}.
\newblock


\bibitem[\protect\citeauthoryear{Potluri, Hamidouche, Venkatesh, Bureddy, and Panda}{Potluri et~al\mbox{.}}{2013}]%
        {potluri2013efficient}
\bibfield{author}{\bibinfo{person}{Sreeram Potluri}, \bibinfo{person}{Khaled Hamidouche}, \bibinfo{person}{Akshay Venkatesh}, \bibinfo{person}{Devendar Bureddy}, {and} \bibinfo{person}{Dhabaleswar~K Panda}.} \bibinfo{year}{2013}\natexlab{}.
\newblock \showarticletitle{Efficient inter-node MPI communication using GPUDirect RDMA for InfiniBand clusters with NVIDIA GPUs}. In \bibinfo{booktitle}{{\em 2013 42nd International Conference on Parallel Processing}}. IEEE, \bibinfo{pages}{80--89}.
\newblock


\bibitem[\protect\citeauthoryear{PyTorch}{PyTorch}{2024a}]%
        {cpp-ext}
\bibfield{author}{\bibinfo{person}{PyTorch}.} \bibinfo{year}{2024}\natexlab{a}.
\newblock \bibinfo{title}{Custom C++ and CUDA Extensions}.
\newblock   (\bibinfo{year}{2024}).
\newblock
\newblock
\shownote{\url{https://pytorch.org/tutorials/advanced/cpp_extension.html}.}


\bibitem[\protect\citeauthoryear{PyTorch}{PyTorch}{2024b}]%
        {third-party-backend}
\bibfield{author}{\bibinfo{person}{PyTorch}.} \bibinfo{year}{2024}\natexlab{b}.
\newblock \bibinfo{title}{Third-party backends}.
\newblock   (\bibinfo{year}{2024}).
\newblock
\newblock
\shownote{\url{https://pytorch.org/docs/stable/distributed.html\#third-party-backends}.}


\bibitem[\protect\citeauthoryear{Rajbhandari, Rasley, Ruwase, and He}{Rajbhandari et~al\mbox{.}}{2020}]%
        {zero}
\bibfield{author}{\bibinfo{person}{Samyam Rajbhandari}, \bibinfo{person}{Jeff Rasley}, \bibinfo{person}{Olatunji Ruwase}, {and} \bibinfo{person}{Yuxiong He}.} \bibinfo{year}{2020}\natexlab{}.
\newblock \bibinfo{title}{ZeRO: Memory Optimizations Toward Training Trillion Parameter Models}.
\newblock   (\bibinfo{year}{2020}).
\newblock
\showeprint[arxiv]{cs.LG/1910.02054}
\showURL{%
\url{https://arxiv.org/abs/1910.02054}}


\bibitem[\protect\citeauthoryear{Ren, Zhou, Meng, Huang, Wang, Wang, Li, Zhang, Podolskiy, Arshinov, Bout, Piontkovskaya, Wei, Jiang, Su, Liu, and Yao}{Ren et~al\mbox{.}}{2023}]%
        {pangu}
\bibfield{author}{\bibinfo{person}{Xiaozhe Ren}, \bibinfo{person}{Pingyi Zhou}, \bibinfo{person}{Xinfan Meng}, \bibinfo{person}{Xinjing Huang}, \bibinfo{person}{Yadao Wang}, \bibinfo{person}{Weichao Wang}, \bibinfo{person}{Pengfei Li}, \bibinfo{person}{Xiaoda Zhang}, \bibinfo{person}{Alexander Podolskiy}, \bibinfo{person}{Grigory Arshinov}, \bibinfo{person}{Andrey Bout}, \bibinfo{person}{Irina Piontkovskaya}, \bibinfo{person}{Jiansheng Wei}, \bibinfo{person}{Xin Jiang}, \bibinfo{person}{Teng Su}, \bibinfo{person}{Qun Liu}, {and} \bibinfo{person}{Jun Yao}.} \bibinfo{year}{2023}\natexlab{}.
\newblock \bibinfo{title}{{PanGu-$\Sigma$: Towards Trillion Parameter Language Model with Sparse Heterogeneous Computing}}.
\newblock   (\bibinfo{year}{2023}).
\newblock
\showeprint[arxiv]{cs.CL/2303.10845}
\showURL{%
\url{https://arxiv.org/abs/2303.10845}}


\bibitem[\protect\citeauthoryear{Shah, Chidambaram, Cowan, Maleki, Musuvathi, Mytkowicz, Nelson, Saarikivi, and Singh}{Shah et~al\mbox{.}}{2023}]%
        {shah2023taccl}
\bibfield{author}{\bibinfo{person}{Aashaka Shah}, \bibinfo{person}{Vijay Chidambaram}, \bibinfo{person}{Meghan Cowan}, \bibinfo{person}{Saeed Maleki}, \bibinfo{person}{Madan Musuvathi}, \bibinfo{person}{Todd Mytkowicz}, \bibinfo{person}{Jacob Nelson}, \bibinfo{person}{Olli Saarikivi}, {and} \bibinfo{person}{Rachee Singh}.} \bibinfo{year}{2023}\natexlab{}.
\newblock \showarticletitle{$\{$TACCL$\}$: Guiding Collective Algorithm Synthesis using Communication Sketches}. In \bibinfo{booktitle}{{\em 20th USENIX Symposium on Networked Systems Design and Implementation (NSDI 23)}}. \bibinfo{pages}{593--612}.
\newblock


\bibitem[\protect\citeauthoryear{Shamis, Venkata, Lopez, Baker, Hernandez, Itigin, Dubman, Shainer, Graham, Liss, Shahar, Potluri, Rossetti, Becker, Poole, Lamb, Kumar, Stunkel, Bosilca, and Bouteiller}{Shamis et~al\mbox{.}}{2015}]%
        {ucx}
\bibfield{author}{\bibinfo{person}{Pavel Shamis}, \bibinfo{person}{Manjunath~Gorentla Venkata}, \bibinfo{person}{M.~Graham Lopez}, \bibinfo{person}{Matthew~B. Baker}, \bibinfo{person}{Oscar Hernandez}, \bibinfo{person}{Yossi Itigin}, \bibinfo{person}{Mike Dubman}, \bibinfo{person}{Gilad Shainer}, \bibinfo{person}{Richard~L. Graham}, \bibinfo{person}{Liran Liss}, \bibinfo{person}{Yiftah Shahar}, \bibinfo{person}{Sreeram Potluri}, \bibinfo{person}{Davide Rossetti}, \bibinfo{person}{Donald Becker}, \bibinfo{person}{Duncan Poole}, \bibinfo{person}{Christopher Lamb}, \bibinfo{person}{Sameer Kumar}, \bibinfo{person}{Craig Stunkel}, \bibinfo{person}{George Bosilca}, {and} \bibinfo{person}{Aurelien Bouteiller}.} \bibinfo{year}{2015}\natexlab{}.
\newblock \showarticletitle{UCX: An Open Source Framework for HPC Network APIs and Beyond}. In \bibinfo{booktitle}{{\em 2015 IEEE 23rd Annual Symposium on High-Performance Interconnects}}.
\newblock


\bibitem[\protect\citeauthoryear{Shoeybi, Patwary, Puri, LeGresley, Casper, and Catanzaro}{Shoeybi et~al\mbox{.}}{2020}]%
        {Megatron-LM}
\bibfield{author}{\bibinfo{person}{Mohammad Shoeybi}, \bibinfo{person}{Mostofa Patwary}, \bibinfo{person}{Raul Puri}, \bibinfo{person}{Patrick LeGresley}, \bibinfo{person}{Jared Casper}, {and} \bibinfo{person}{Bryan Catanzaro}.} \bibinfo{year}{2020}\natexlab{}.
\newblock \bibinfo{title}{Megatron-LM: Training Multi-Billion Parameter Language Models Using Model Parallelism}.
\newblock   (\bibinfo{year}{2020}).
\newblock
\showeprint[arxiv]{cs.CL/1909.08053}
\showURL{%
\url{https://arxiv.org/abs/1909.08053}}


\bibitem[\protect\citeauthoryear{Tillet, Kung, and Cox}{Tillet et~al\mbox{.}}{2019}]%
        {tillet2019triton}
\bibfield{author}{\bibinfo{person}{Philippe Tillet}, \bibinfo{person}{Hsiang-Tsung Kung}, {and} \bibinfo{person}{David Cox}.} \bibinfo{year}{2019}\natexlab{}.
\newblock \showarticletitle{Triton: an intermediate language and compiler for tiled neural network computations}. In \bibinfo{booktitle}{{\em Proceedings of the 3rd ACM SIGPLAN International Workshop on Machine Learning and Programming Languages}}. \bibinfo{pages}{10--19}.
\newblock


\bibitem[\protect\citeauthoryear{Um, Oh, Kang, Lee, Kim, Kim, Kim, Muzzammil, and Jeon}{Um et~al\mbox{.}}{2024}]%
        {metis}
\bibfield{author}{\bibinfo{person}{Taegeon Um}, \bibinfo{person}{Byungsoo Oh}, \bibinfo{person}{Minyoung Kang}, \bibinfo{person}{Woo-Yeon Lee}, \bibinfo{person}{Goeun Kim}, \bibinfo{person}{Dongseob Kim}, \bibinfo{person}{Youngtaek Kim}, \bibinfo{person}{Mohd Muzzammil}, {and} \bibinfo{person}{Myeongjae Jeon}.} \bibinfo{year}{2024}\natexlab{}.
\newblock \showarticletitle{Metis: Fast Automatic Distributed Training on Heterogeneous {GPUs}}. In \bibinfo{booktitle}{{\em 2024 USENIX Annual Technical Conference (USENIX ATC 24)}}. \bibinfo{publisher}{USENIX Association}, \bibinfo{address}{Santa Clara, CA}, \bibinfo{pages}{563--578}.
\newblock
\showISBNx{978-1-939133-41-0}
\showURL{%
\url{https://www.usenix.org/conference/atc24/presentation/um}}


\bibitem[\protect\citeauthoryear{Venkata, Petrov, Lebedev, Bureddy, Aderholdt, Ladd, Bloch, Dubman, and Shainer}{Venkata et~al\mbox{.}}{2024}]%
        {DBLP:conf/hoti/VenkataPLBALBDS24}
\bibfield{author}{\bibinfo{person}{Manjunath~Gorentla Venkata}, \bibinfo{person}{Valentine Petrov}, \bibinfo{person}{Sergey Lebedev}, \bibinfo{person}{Devendar Bureddy}, \bibinfo{person}{Ferrol Aderholdt}, \bibinfo{person}{Joshua Ladd}, \bibinfo{person}{Gil Bloch}, \bibinfo{person}{Mike Dubman}, {and} \bibinfo{person}{Gilad Shainer}.} \bibinfo{year}{2024}\natexlab{}.
\newblock \showarticletitle{Unified Collective Communication {(UCC):} An Unified Library for CPU, GPU, and {DPU} Collectives}. In \bibinfo{booktitle}{{\em {IEEE} Symposium on High-Performance Interconnects, {HOTI} 2024, Albuquerque, NM, USA, August 21-23, 2024}}. \bibinfo{publisher}{{IEEE}}, \bibinfo{pages}{37--46}.
\newblock
\showDOI{%
\url{https://doi.org/10.1109/HOTI63208.2024.00018}}


\bibitem[\protect\citeauthoryear{Yim, Song, Choi, Lee, Jung, Jang, and Lee}{Yim et~al\mbox{.}}{2024}]%
        {pipette}
\bibfield{author}{\bibinfo{person}{Jinkyu Yim}, \bibinfo{person}{Jaeyong Song}, \bibinfo{person}{Yerim Choi}, \bibinfo{person}{Jaebeen Lee}, \bibinfo{person}{Jaewon Jung}, \bibinfo{person}{Hongsun Jang}, {and} \bibinfo{person}{Jinho Lee}.} \bibinfo{year}{2024}\natexlab{}.
\newblock \showarticletitle{Pipette: Automatic Fine-Grained Large Language Model Training Configurator for Real-World Clusters}. In \bibinfo{booktitle}{{\em 2024 Design, Automation and Test in Europe Conference and Exhibition, DATE 2024 - Proceedings}} {\em (\bibinfo{series}{Proceedings -Design, Automation and Test in Europe, DATE})}. \bibinfo{publisher}{Institute of Electrical and Electronics Engineers Inc.}, \bibinfo{address}{United States}.
\newblock
\newblock
\shownote{Publisher Copyright: {\textcopyright} 2024 EDAA.; 2024 Design, Automation and Test in Europe Conference and Exhibition, DATE 2024 ; Conference date: 25-03-2024 Through 27-03-2024.}


\bibitem[\protect\citeauthoryear{Zhao, Maleki, Shah, Yang, Pourreza, and Krishnamurthy}{Zhao et~al\mbox{.}}{2024}]%
        {zhao2024forestcoll}
\bibfield{author}{\bibinfo{person}{Liangyu Zhao}, \bibinfo{person}{Saeed Maleki}, \bibinfo{person}{Aashaka Shah}, \bibinfo{person}{Ziyue Yang}, \bibinfo{person}{Hossein Pourreza}, {and} \bibinfo{person}{Arvind Krishnamurthy}.} \bibinfo{year}{2024}\natexlab{}.
\newblock \showarticletitle{Forestcoll: Efficient collective communications on heterogeneous network fabrics}.
\newblock \bibinfo{journal}{{\em arXiv preprint arXiv:2402.06787\/}} (\bibinfo{year}{2024}).
\newblock


\bibitem[\protect\citeauthoryear{Zheng, Bao, Hou, Zheng, Fang, Huang, Li, Duanmu, Chen, Xu, et~al\mbox{.}}{Zheng et~al\mbox{.}}{2025}]%
        {zheng2025triton}
\bibfield{author}{\bibinfo{person}{Size Zheng}, \bibinfo{person}{Wenlei Bao}, \bibinfo{person}{Qi Hou}, \bibinfo{person}{Xuegui Zheng}, \bibinfo{person}{Jin Fang}, \bibinfo{person}{Chenhui Huang}, \bibinfo{person}{Tianqi Li}, \bibinfo{person}{Haojie Duanmu}, \bibinfo{person}{Renze Chen}, \bibinfo{person}{Ruifan Xu}, {et~al\mbox{.}}} \bibinfo{year}{2025}\natexlab{}.
\newblock \showarticletitle{Triton-distributed: Programming Overlapping Kernels on Distributed AI Systems with the Triton Compiler}.
\newblock \bibinfo{journal}{{\em arXiv preprint arXiv:2504.19442\/}} (\bibinfo{year}{2025}).
\newblock


\bibitem[\protect\citeauthoryear{Zhong, Liu, Chen, Hu, Zhu, Liu, Jin, and Zhang}{Zhong et~al\mbox{.}}{2024}]%
        {zhong2024distserve}
\bibfield{author}{\bibinfo{person}{Yinmin Zhong}, \bibinfo{person}{Shengyu Liu}, \bibinfo{person}{Junda Chen}, \bibinfo{person}{Jianbo Hu}, \bibinfo{person}{Yibo Zhu}, \bibinfo{person}{Xuanzhe Liu}, \bibinfo{person}{Xin Jin}, {and} \bibinfo{person}{Hao Zhang}.} \bibinfo{year}{2024}\natexlab{}.
\newblock \showarticletitle{$\{$DistServe$\}$: Disaggregating prefill and decoding for goodput-optimized large language model serving}. In \bibinfo{booktitle}{{\em 18th USENIX Symposium on Operating Systems Design and Implementation (OSDI 24)}}. \bibinfo{pages}{193--210}.
\newblock


\end{thebibliography}
